\documentclass{lmcs} 
\pdfoutput=1
\usepackage[utf8]{inputenc}

\usepackage{lastpage}
\lmcsdoi{22}{1}{29}
\lmcsheading{}{\pageref{LastPage}}{}{}%
{Mar.~31,~2025}{Mar.~31,~2026}{}

\keywords{reactive synthesis, computational complexity, formal methods, controller synthesis}

\usepackage{amssymb, amsmath, mathrsfs}
\usepackage{tikz}
\usetikzlibrary{shapes,arrows,positioning, cd, calc, matrix}
\usepackage{csquotes}
\usepackage[draft, color=green!40]{todonotes}
\usepackage{caption, subcaption}
\usepackage{booktabs}   
\usepackage{multirow}   
\usepackage{hyperref}

\usetikzlibrary{shapes.geometric}
\newcommand{\none}{\text -}
\newcommand{\Situation}[2]{$\underset{(#2)}{{#1}}$}
\tikzset{ ego/.style = {draw, circle, minimum size=0.8cm, inner sep=0pt}, alter/.style = {shape aspect=1,diamond,draw, minimum size=0.8cm, inner sep=-0.5pt}, winnable/.style = {fill=gray!50}, EconstViolated/.style = {double}, AconstViolated/.style = {dashed}, ddd/.style = {minimum size=0.2cm}}
\newcommand{\SH}[2]{$s_{#1},H_{#2}$}

\newcommand{\E}{EGO} 
\newcommand{\A}{ALTER}
\newcommand{\N}{\mathbb{N}}

\DeclareSymbolFont{largesymbolsA}{U}{txexa}{m}{n}
\DeclareMathSymbol{\varprod}{\mathop}{largesymbolsA}{16}

\theoremstyle{plain} 

\def\eg{{\em e.g.}}

\def\ie{{\em i.e.}}

\begin{document}
		
		\title[Window Counting Constraints for the Synthesis of Reactive Systems]{Towards the Usage of Window Counting Constraints \texorpdfstring{\\}{} in the Synthesis of Reactive Systems to Reduce State Space Explosion\rsuper*}
		\titlecomment{{\lsuper*} This is an extended version of \cite{FeekenFraenzle2024} presented at the GandALF 2024 conference.}
		
		\author[L.~Feeken]{Linda Feeken\lmcsorcid{0009-0003-7336-0859}}[a]
		\author[M.~Fränzle]{Martin Fränzle\lmcsorcid{0000-0002-9138-8340}}[b]
		
		\address{German Aerospace Center (DLR), Oldenburg, Germany}	
		\email{linda.feeken@dlr.de}  
		
		\address{Carl von Ossietzky Universität Oldenburg, Oldenburg, Germany}	
		
		
		
		
		
		\begin{abstract}
		The synthesis of reactive systems aims for the automated construction of strategies for systems that interact with their environment. 
		Whereas the synthesis approach has the potential to change the development of reactive systems significantly due to the avoidance of manual implementation, it still suffers from a lack of efficient synthesis algorithms for many application scenarios. The translation of the system specification into an automaton that allows for strategy construction (if a winning strategy exists) is nonelementary in the length of the specification in S1S and doubly exponential for LTL, raising the need of highly specialized algorithms. 
		In this article, we present an approach on how to reduce this state space explosion in the construction of this automaton by exploiting a monotonicity property of specifications. 
		For this, we introduce window counting constraints that allow for step-wise refinement or abstraction of specifications.
		In an iterative synthesis procedure, those window counting constraints are used to construct automata representing over- or under-approximations (depending on the counting constraint) of constraint-compliant behavior. Analysis results on winning regions of previous iterations are used to reduce the size of the next automaton, leading to an overall reduction of the state space explosion extent.
		We present the implementation results of the iterated synthesis for a zero-sum game setting as proof of concept. Furthermore, we discuss the current limitations of the approach in a zero-sum setting and sketch future work in non-zero-sum settings. 
		\end{abstract}
		
		\maketitle
		
		\section*{Introduction}\label{sec:intro}
		
		The automated translation of a system specification into its implementation is one of the most challenging problems in formal methods. Providing correct-by-construction artifacts, such synthesis offers great potential in the development of new systems by significantly reducing the need for manual work in the engineering process.  
		In this article, we focus on synthesis for reactive systems, i.e. systems that are influenced by and interact with their environment. This interaction can be modeled as a game, in which the system tries to play according to its specification, whereas the moves of the environment can potentially impede the system from reaching its goal. Since the interaction between system and environment is typically of long-lasting nature without predefined end date, the game is infinite in the sense that a play of the game has infinite duration, while the arena, modeled as a graph, has finitely many states. The players play by moving a token from one state of the arena to the next. The player whose turn it is decides which of the outgoing transitions of the current state is chosen. A well-known type of game is the safety game: The system wins a play if it can avoid to reach predefined unsafe states. Otherwise, the environment wins. A player has a winning strategy, if it wins against all possible behavior of the other player. For two-player safety games on finite graphs, there always exists a winning strategy for one of the players and this winning strategy can be computed \cite{buchi1967solving}, \cite{thomas1995}. 
		However, the efficient computation of winning strategies (not only in the case of safety games) is still an open challenge in the synthesis of reactive systems. 
		A common synthesis approach is to first generate a deterministic word automaton as game graph from specifications written as Linear Temporal Logic (LTL) formulae. Then, a strategy that is winning in the game is calculated. By construction, the strategy automatically satisfies the specification. Unfortunately, the construction of the deterministic word automaton leads to an automaton with a number of states that is doubly exponential in the length of the specification \cite{PnueliR89}, making the whole strategy synthesis unfeasible for many applications. For avoiding the most expensive part of the synthesis procedure, there exist synthesis algorithms that start with a subset of the specification language LTL, such that it is possible to construct the game graph in a more efficient way. One example for that is the usage of the LTL subclass Generalized Reactivity(1) (GR(1)), which allows to construct and solve the game in time $O(N^3)$ with $N$ being the size of the state space \cite{piterman2006synthesis}. While GR(1) is expressive enough for the specification of many systems \cite{maoz2015gr}, some specifications that do not fall into GR(1) remain unconsidered. For example, Maoz and Ringert mention the consideration of synthesis with counting patterns as future work in \cite{maoz2015gr}, but to the best of our knowledge, this has not been addressed yet.
		
		In this article, we deal with the request for heuristically efficient synthesis for some types of counting patterns as part of the system specification and present the idea of recursive synthesis for such games. 
		The translation of the full system specification into a deterministic automaton as game arena is out of scope. We already start with a deterministic automaton accompanied with a set of counting patterns. The automaton can be obtained from a specification by any translation procedure (without encoding the counting patterns as well) or can arise naturally like it often occurs in the robotics domain, where the graph encodes the position of an automated system in specified areas like a factory floor.
		We call the considered counting patterns \enquote{window counting constraints}. These are of the form \\
		\centerline{\textit{\enquote{The system plays action $act$ at least $k$ times out of $l$ of its own moves.}}} \\
		with parameters $k, l\in \mathbb{N}, \, k\leq l$ and $act$ encoding an action controllable by the system. 
		The \enquote{at least} can also be replaced by \enquote{at most}. 
		Such constraints arise naturally when the desired behavior of systems includes reoccurring elements. For example, an automated guided vehicle on a factory floor might need to charge its battery in at least two out of ten moves to avoid facing depleted batteries. 
		By replacing \enquote{system} in the window counting constraint by \enquote{environment}, the constraint introduces rationality of the environment, since the environment will not primary behave antagonistically to the system in plays of the game, but will behave according to its own constraints. From the synthesis point of view, rationality of the environment implies information for the system on the environment, reducing the space of situations the system could face in a play of the game.
		The term \enquote{window} in the constraint-type name  emphasizes the relation of those specifications to sliding windows in data stream monitoring \cite{PatroumpasS06}. For the sake of readability, we also call them counting constraints for short.\\
		
		We avoid the direct full translation of the specifications in form of counting constraints into a graph and instead focus on the following two observations:
		(1) It is possible to influence how hard it is to satisfy a counting constraint by varying the parameters $k$, $l$ in the counting constraints. More precisely, the (non-)existence of a strategy that fulfills the specification in a game with a set of counting constraints allows to make statements about the (non-)existence of such a strategy in a game with a set of counting constraints with varied parameters. \linebreak
		(2) The values in the counting constraints influence the scale of the game graph that encodes all information given by the constraints. The larger $k$ and $l$, the exponentially larger in the (unary) values of $k$ and $l$ is the graph, if the memory on the last moves is encoded into the states of the graph. Consequently, the values influence how much computational power and/or memory is needed in order to synthesize a winning strategy.
		
		Combining these observations, the presented approach can be summarized as follows: 
		Consider a two-player game graph and some specifications in the form of counting constraints. For solving the synthesis problem of finding a strategy for the system, such that the counting constraints are fulfilled,  
		start with a subset of counting constraints that result in a small game graph or a trivially winnable game. Calculate winning strategies (if existent) and check what the (non-)existence of a winning strategy means for a game with refined/relaxed (depending on the constraints) constraints. 
		This information gives reliable hints on which parts of the game with adapted values in the counting constraints are worth to investigate in the next iteration step and which parts of the game graph can then be neglected, leading to a reduction in the state space. With each iteration, the set of considered counting constraints converges to the game of interest. 
		Although the size of the game graphs may increase in each iteration, the gained state space reduction leads to a synthesis algorithm heuristically more efficient than when considering the game of interest as a whole from the beginning.
		If the original game without window counting constraints is a reachability, safety, Büchi, co-Büchi or parity-game, then the synthesis task in each iteration can be done by any synthesis algorithm fitting to the game class. 
		This approach is not improving the complexity class of the synthesis problem, but shall serve as suggestion on how to improve existing synthesis algorithms heuristically for games that use counting constraints. 

		This work is an extension of \cite{FeekenFraenzle2024}. The extended version complements the content of the original one in two main areas: First, the synthesis approach in \cite{FeekenFraenzle2024} only considered safety games, while this work extends the game types to reachability, Büchi, co-Büchi and parity games. Second, this work adds the possibility of iterating over constraints of both the form \enquote{The system plays action $act$ \textit{at least} $k$ times out of $l$ of its own moves.} and \enquote{The system plays action $act$ \textit{at most} $k$ times out of $l$ of its own moves.} in the same synthesis procedure, while the previous version of the approach considered one of the constraint types to be fixed. Additionally, this article is enriched with more extensive explanations and examples.
		
		This article is structured as follows. 
		In \autoref{sec:relatedWork}, related work in the field of synthesis for reactive systems is presented, focusing on the challenge of constructing efficient algorithms.
		After summarizing concepts and notations required to formulate the game, \autoref{sec:gamesWithCountingConstraints} provides the definition of a game with counting constraints.
		In \autoref{sec:iteratedSynthesis}, we present the idea of incremental synthesis with counting constraints and an algorithm for it. An example and runtime results of the algorithm based on a non-optimized implementation are given in \autoref{sec:experiments}. Current limitations and chances of the incremental synthesis approach that guide our future work are collected in \autoref{sec:discussion}.  \autoref{sec:conclusion} concludes the article.
		
		\section{Related Work}\label{sec:relatedWork}
		
		In 1957, Church formulated the Synthesis Problem as finding finite-memory procedures to transform an infinite sequence of input data into an infinite sequence of output data, such that the relation between input and output satisfies given specifications \cite{church1957}, \cite{Thomas09}. Around a decade later, Büchi and Landweber showed the decidability of the problem \cite{buchi1967solving}. However, the algorithmic complexity of synthesis algorithms remains a challenge. 
		The translation of specifications from monadic second-order logic of one successor (S1S) into a Büchi automaton as part of the synthesis procedure is nonelementary in the length of specifications \cite{Stockmeyer74}. This indicates that it is not possible to construct a universally efficient synthesis algorithm that can handle complete S1S specifications.
		For specifications expressible in Linear Temporal Logic (LTL), the problem is 2EXPTIME-complete \cite{PnueliR89}. \\
		Acknowledging the absence of a generally low-complexity synthesis algorithm for arbitrary S1S/LTL specifications, the literature presents three primary approaches \cite{Finkbeiner2016}. 
		The first approach restricts the scope of considered specifications for synthesis to less expressive logics.
		Here, the structure of the considered specifications is used to reduce the synthesis complexity.
		The second one is tackling the internal representation of the problem. Solutions following this approach are often aiming for algorithms with in average good runtime. In this approach, it suffices if most systems can be synthesized with acceptable resources (memory, computational time), while the existence of corner cases with worst-case complexity is accepted. 
		The third approach focuses on the output of the problem, the implementation. The size of the implementation gets restricted such that only small implementations are accepted as solutions of the synthesis problem. The rationale behind this is that small and hence less complicated implementations often exist for applications. Such solutions are often easier (that is, with less computational time) identifiable than bigger (complex) implementations, if it is possible to steer the algorithm towards small solutions. Synthesis algorithms can follow more than one of those approaches.
		
		A well-studied class of specifications for approach (1) is General Reactivity of Rank 1 (GR1),  
		a fragment of LTL for which there are symbolic synthesis algorithms that are polynomial in the size of the state space of the design \cite{piterman2006synthesis}.
		Examples for other specification classes for which efficient solutions of the synthesis problem are investigated are Safety LTL \cite{ZhuTLPV17}, Metric Temporal Logic with a Bounded Horizon \cite{MalerNP07} and Extended Bounded Response LTL \cite{CimattiGGMT20}.
		
		Following approach (2), Kupferman and Vardi developed a synthesis method that does not require the costly determinization of non-deterministic Büchi automata representing the specification \cite{KupfermanV05}, which is the most complex part in many synthesis algorithms. Other synthesis algorithms rely for instance on symbolic synthesis to represent sets of states of a game graph in a compact matter via antichains \cite{FiliotJR09}, \cite{FiliotJR11}, binary decision diagrams \cite{Ehlers10} and LTL fragments \cite{CimattiGGMT20}.
		
		The work by Schewe and Finkbeiner presents a synthesis algorithm that employs bounded synthesis as approach (3). Their method uses translation of LTL specifications into sequences of safety tree automata, in order to constrain the size of the implementation \cite{ScheweF07a}.
		\enquote{Lazy synthesis}, in which an SMT solver is used to construct potential implementations for an incomplete constraint system, extends the system only if required \cite{FinkbeinerJ12}.
		
		The synthesis algorithm presented in this article includes elements of approaches (2) and (3). We avoid the full construction of an automaton representing the specifications by starting with a small specification that is successively enlarged. In each step, the size of the resulting automaton is reduced (if possible). The procedure stops if a winning strategy can already be found in some intermediate step, leading to small solutions. However, it is not possible to restrict the size of the implementation directly as commonly done in bounded synthesis. 
		
		The general idea is inspired by the work of Chen et al. on safety games with delay. In this work, one player only receives information on the moves of the environment with a delay of $k \in \mathbb{N}$ turns. In particular, the player is forced to decide on its next move without having knowledge on the latest k turns of the recent history of the play. The key idea is that a strategy that is not winning for the player for some delay $l\leq k$ will also not be winning in the game with delay $k$. Moreover, moves of the system that are not part of a winning strategy for the game with delay $l$ will also not be part of a winning strategy for the game with larger delay. Hence, such behavior does not need to be considered for larger delays, allowing to reduce the required transition system that encodes the game with a larger delay (pruning step). The synthesis algorithm begins with assuming a delay of zero (no delay), which is the well-known problem of solving a safety game. If a winning strategy for the player exists, the constructed transition system is pruned and the delay is increased by one for the next iteration. If no winning strategy exists in any iteration for delay $l  \leq k$, no winning strategy for delay $k$ exists \cite{chen2018s}. 
		
		\section{Games with Counting Constraints} \label{sec:gamesWithCountingConstraints}
		This section introduces games with counting constraints after repeating standard definitions for two-player games that are needed to formalize the presented game. 
		
		\begin{defi}[Two-player game graph] \label{def:twoPlayerGameGraph}
			A \textbf{two-player finite-state game graph} (or: arena) is of the form $A = (S, s_0, S_{\E}, S_{\A}, \Sigma_{\E}, \Sigma_{\A}, \rightarrow)$ where $S$ is a finite (non-empty) set of states, $S_{\E}$, $S_{\A}$ define a partition of $S$, $s_0 \in S_{\E}$ is the initial state, $\Sigma_{\E}$ is a finite alphabet of actions for player $\E$,  $\Sigma_{\A}$ is a finite alphabet of actions for player $\A$ and 
			$\rightarrow \subseteq \bigcup_{p \in \{\E, \A}\} S_p \times \mathcal{P}(\Sigma_p) \times S_{\overline{p}}$
			is a set of labeled transitions satisfying the following two conditions, where $\mathcal{P}(M)$ denotes the power set of a set $M$, $\overline{\E} = \A$ and $\overline{\A} = \E$:
			\begin{itemize}
				\item Absence of deadlock: For each $s \in S$ there exists $\sigma \in \mathcal{P}(\Sigma_{\E}) \cup \mathcal{P}(\Sigma_{\A})$ and $s' \in S$, such that $(s,\sigma,s') \in \rightarrow$.
				\item Determinism of moves: For all $(s,\sigma,s'), (s,\sigma,s'') \in \rightarrow$ holds $s' = s''$.  
			\end{itemize}
		\end{defi}
		
		Such a game graph, also referred to as \enquote{arena}, encodes a game between the two players $\E$ and $\A$. For $p \in \{\E, \A\}$ the set of states $S_p$ contains the states where it is the turn of player $p$ to perform an action, also called \enquote{$p$ controls $s$}. 
		The game is \enquote{turn-based}, i.e.\ the two players alternate between choosing one of the possible actions. Since the game graph does not contain deadlocks, it results in an infinite sequence of states and actions, called an infinite play.
		
		\begin{defi}[Infinite play, prefix]
			Let  $A = (S, s_0, S_{\E}, S_{\A}, \Sigma_{\E}, \Sigma_{\A}, \rightarrow)$ be a two-player game graph. An \textbf{infinite play} on $A$ is an infinite sequence $\pi = (\pi_i \sigma_i)_{i\in \N_0} = \pi_0 \sigma_0 \pi_1 \sigma_1 \dots$ with $\pi_0 = s_0$ and $\pi_i \sigma_i \pi_{i+1} \in \rightarrow$ for all $i \in \N_0$. $\Pi(A)$ denotes the set of all infinite plays on $A$. 
			A finite \textbf{prefix} of an infinite play $\pi$ is $h = (\pi_i \sigma_i )_{0 \leq i \leq n} = \pi_0 \sigma_0 \pi_1 \sigma_1 \dots \sigma_{n-1}\pi_n$ (for some $n \in \mathbb{N}_0$). It ends in $Tail(h) = \pi_n$, called its \textbf{tail}. It \emph{belongs} to $\E$ if $\pi_n \in S_{\E}$ and to $\A$ otherwise. The set of prefixes of $\E$ (respectively, $\A$) is denoted $\mathrm{Pref}_{\E}(A)$ (resp., $\mathrm{Pref}_{\A}(A)$), and the set of prefixes is $\mathrm{Pref} = \mathrm{Pref}_{\E} \uplus \mathrm{Pref}_{\A}$.
		\end{defi}

		Since the two players alternate in choosing their actions, we can also interpret a play $\pi \in \Pi(A)$ as interleaving $\pi = u \otimes v := (u_i v_i)_{i\in \N_0}$ of unique sequences $u =(u_i)_{i	\in \N_0}\in (S_{\E} \times \mathcal{P}(\Sigma_{\E}))^{\omega}$, $v =(v_i)_{i	\in \N_0}\in (S_{\A} \times \mathcal{P}(\Sigma_{\A}))^{\omega}$. A prefix can we interpreted likewise.
		If the played letters are not of relevance in a play, we also denote plays solely by the visited situations, \ie, $\pi \in S^{\omega}$.
		In such an infinite play, the two players play against (or in case of collaborative games: with) each other. Players can have strategies that determine how they react in each step of the play.
		
		\begin{defi}[Strategy]
			Let  $A = (S, s_0, S_{\E}, S_{\A}, \Sigma_{\E}, \Sigma_{\A}, \rightarrow)$ be a two-player game graph.
			\begin{itemize}
				\item A \textbf{strategy} for a player $p \in \{\E, \A\}$ in the game graph $A$ is a mapping $\varphi: \, \mathrm{Pref}_{p}(A) \longrightarrow \mathcal{P}(\Sigma_{p})$, such that for all prefixes $h \in \mathrm{Pref}_p(A)$ and all $\sigma \in \varphi(h)$ there exist a state $s \in S \setminus S_p$ and a transition $(Tail(h), \sigma, s) \in \rightarrow$. 
				\item A play $\pi$ on a game graph $A$ is an \textbf{outcome} of a strategy $\varphi$ of $p \in \{\E, \A\}$ in the game graph $A$ when player $p$ follows the strategy $\varphi$ and the other player plays arbitrarily, i.e.: for all prefixes $h \pi_n \sigma_n$ (for some $\pi_n \in S$ and $\sigma_n \in \mathcal{P}(\Sigma)$) that are prefixes of $\pi$, if $h \in \mathrm{Pref}_p(A)$, then $\sigma_n = \varphi(h \pi_n)$ (and if $h \in \mathrm{Pref}_{\overline{p}}(A)$, no condition is imposed). The set of \textbf{outcomes} of $\varphi$ on game graph $A$ is denoted $O(A, \varphi)$.
				\item A strategy $\varphi$ for player $p \in \{\E, \A\}$ in the game graph $A$ is \textbf{positional}, if the players next move only depends on the current state. In particular, a positional strategy requires no memory on the prefix of a run except for the current state. Formally: $\varphi$ is positional $:\Leftrightarrow \forall \pi , \omega \in \mathrm{Pref}_p(A): Tail(\pi) = Tail(\omega) \Rightarrow \varphi(\pi) = \varphi(\omega)$.
			\end{itemize}
		\end{defi}
		
		A game $G = (A, \text{Win})$ consists of a game graph $A$ and a winning condition \text{Win} like a reachability, safety, Büchi, co-Büchi or parity condition. The winning condition consist of a set of (infinite) sequences of states of $G$.
		$\E$ wins if it has a strategy $\varphi$ that guarantees $O(A, \varphi) \subseteq \text{Win}$. The environment, on the other hand, wins if a play not in \text{Win} is played. Hence, each play of such a two-player game with $\omega$-regular winning condition always has exactly one winner and one loser. Games with this property are called zero sum games. \\
		In this article, safety games are used in all examples. Note that the presented approach is applicable to other winning conditions like the ones below. Required winning condition-specific modifications are described in the explanation of the approach (Section~\ref{sec:iteratedSynthesis}).
		
		\begin{defi}[Winning conditions]
			Let $A$ be a game graph with states $S$ and $R \subseteq S$.
			For a play $\pi$ on $A$ denote the set of states of $A$ that are visited at least once in $\pi$ with $Occ(\pi)$ and the set of states that are visited infinitely often in $\pi$ as $Inf(\pi)$.
			$G=(A, \text{Win})$ becomes a game with any of the following winning conditions:
			
			\begin{itemize}
				\item The \textbf{safety} winning condition is defined as $\text{Win} = \mathtt{Safe}(R) := \{\pi \in S^{\omega} \, | \, Occ(\pi) \subseteq R\}$. In a safety game, player $\E$ wins a play, if only \enquote{safe} states $s\in R$ are visited. States $s \in S\setminus R$ are called unsafe states.
				\item The \textbf{reachability} winning condition is defined as $\text{Win} = \mathtt{Reach}(R) := \{\pi \in S^{\omega} \, | \, Occ(\pi) \cap R \neq \emptyset \}$. In a reachability game, player $\E$ wins a play, if any state $s\in R$ is visited at least once in the play.
				\item The \textbf{Büchi} winning condition is defined as $\text{Win} = \mathtt{B\ddot{u}chi}(R) := \{\pi \in S^{\omega} \, | \, Inf(\pi) \cap R \neq \emptyset\}$. In a Büchi game, player $\E$ wins a play, if there is any state in $R$ that is visited infinitely often during the play.
				\item The \textbf{co-Büchi} winning condition is defined as $\text{Win} = \mathtt{coB\ddot{u}chi}(R) := \{\pi \in S^{\omega} \, | \, Inf(\pi) \subseteq R\}$. In a co-Büchi game, player $\E$ wins a play, if the only states that are visited infinitely often during the play are elements of $R$.
				\item Define a coloring $\Omega \colon S \to \N_0$. The \textbf{parity} winning condition is defined as $\text{Win} = \mathtt{Parity}(\Omega) := \{\pi \in S^{\omega} \, | \, min( Inf(\Omega(\pi_i)_{i \in \N_0})) \text{ is even}\}$. In a parity game, player $\E$ wins a play, if the smallest color that is shown infinitely often is even.
			\end{itemize}
		\end{defi}
		
		A game with any of the above winning conditions is positionally determined, that is, for any state of the game, one of the players has a positional winning strategy.\\
		We are now introducing window counting constraints as a mean to encode reoccurring behavior of a player with limits on which action can be selected how often in each snippet (or: window) of a play of a given length. Each window counting constraint belongs to a player (either $\E$ or $\A$) and the action is a formula over the alphabet of the player.
		
		\begin{defi}[Window Counting Constraints]\label{def:windowCountingConstraints}
			Let $A$ be a game graph with the two players $\E$ and $\A$. Let $p \in {\E, \A}$.
			Let $\pi = u^{\E} \otimes u^{\A}  = (u^{\E}_i u^{\A}_i)_{i\in \mathbb{N}_0}$ be a play on $G$.
			Denote with $a$ a propositional logic formula over the alphabet $\Sigma_{p}$ and $k, l \in \mathbb{N}$ with $k \leq l$.   
			Define for $i\in \N_0$ and $u^{p}_i = (s,\sigma)\in S_p \times \mathcal{P}(\Sigma_p)$ $u^p_i \models a : \Leftrightarrow \land_{b \in \sigma} b \models a$.
			\begin{enumerate}
				\item \textbf{$CC_{max}(p, a, k, l)$} is defined as the abbreviation for \enquote{The player $p$ plays $a$ at most $k$ times out of $l$ of its own turns.}. 
				$CC_{max}(p, a, k, l)$ is satisfied on $\pi$, if for all $i\in \mathbb{N}_0$ holds $|\{m \in \mathbb{N}_0 \, | \, u^p_m \models a,\, i\leq m \leq i+l\}|\leq k$.
				\item \textbf{ $CC_{min}(p, a, k, l)$} is defined as the abbreviation for \enquote{The player $p$ plays $a$ at least $k$ times out of $l$ of its own turns.}. 
				$CC_{min}(p, a, k, l)$ is satisfied on $\pi$, if for all $i\in \mathbb{N}_0$ holds $|\{m \in \mathbb{N}_0 \, | \, u^p_m \models a,\, i\leq m \leq i+l\}|\geq k$.	
			\end{enumerate}
			A prefix of a play on $A$ satisfies a counting constraint, if it can be complemented to an infinite play that satisfies the counting constraint in any way (in particular, the extended prefix does not need to be a play on $A$).
			The parameter $l$ is called the length of a counting constraint.
		\end{defi}
		
		The definition of satisfying a counting constraint for a play  is extended canonically to satisfying a set of counting constraints and counting constraints being satisfied on a strategy.
		
		\begin{exa} \label{exa:countingConstraints}
			Figure~\ref{fig:example_countingConstraints} shows a small game graph with $\Sigma_{\E} = \{x, y\}$ and $\Sigma_{\A} = \{a, b, c\}$ being the alphabets of the two players $\E$ and $\A$. Consider the window counting constraints 
			\begin{itemize}
				\item \enquote{The player $\E$ plays $x$ at least $2$ times out of $4$ of its own turns.},
				\item \enquote{The player $\E$ plays $y$ at most $2$ times out of $3$ of its own turns.},
				\item \enquote{The player $\E$ plays $y$ at least $2$ times out of $5$ of its own turns.} and
				\item \enquote{The player $\A$ plays $a \vee b$ at least $1$ time out of $2$ of its own turns.}. 
			\end{itemize}   
			The play $\pi = (1, \{y\}, 2, \{a\}, 3, \{x\}, 2, \{a\}, 3, \{x,y\}, 4, \{a\})^{\omega}$ on the game graph satisfies all of the four window counting constraints, since each \enquote{sliding window} shows constraint satisfaction. Figure~\ref{fig:prefix_fulfilling_countstraint} illustrates the constraint fulfillment for the first two constraints. \\
			The prefix $pre = (1, \{x\}, 4, \{a\}, 1, \{x\}, 4, \{x\},1, \{x\}, 4)$ of another play on the shown graph is not satisfying \enquote{The player $\E$ plays $y$ at least $2$ times out of $5$ of its own turns.}, since no matter what $\E$ does in its next move (even if it would be allowed to change the game graph for the next turns), it is not possible to extend $pre$ to a play that satisfies the constraint.
		\end{exa}
		
		\begin{figure} 
			\centering
			\begin{tikzpicture}[node distance={16mm}, thick, scale=.6] 
				\node[ego] (1) {$1$}; 
				\node[alter] (2) [right of=1] {$2$}; 
				\node[ego] (3) [below of=2] {$3$};
				\node[alter] (4) [left of=3] {$4$};
				\node[ego] (5) [left of=4] {$5$};
				
				\draw[->] (-1,0) -- (1); 
				\draw[->] (1) -- node[midway, above] {y} (2); 
				\draw[->] (2) to [out=225,in=125]  node[midway, right] {a} (3); 
				\draw[->] (3) to [out=45,in=315]  node[midway, left] {x} (2); 
				\draw[->] (3) to [out=145,in=35]  node[midway, below] {x, y} (4); 
				\draw[->] (4) to [out=-45,in=225]  node[midway, above] {b} (3); 
				\draw[->] (4) to [out=225,in=315]  node[midway, above] {c} (5); 
				\draw[->] (5) to [out=45,in=145]  node[midway, below] {y} (4); 
				\draw[->] (4) to [out=125,in=225]  node[midway, right] {a} (1); 
				\draw[->] (1) to [out=-45,in=55]  node[midway, left] {x} (4);
			\end{tikzpicture} 
			\caption{Small game graph for a two-player game. Circles represent states controlled by $\E$, diamond-shaped states are controlled by $\A$. Transitions are equipped with the actions a player plays when deciding for a transition. For example, $\E$ can choose in state 3 to play $x$ and $y$ in one turn (leading to state 4) or just play $x$ (and not $y$), which leads to state 2.}
			\label{fig:example_countingConstraints}
		\end{figure}
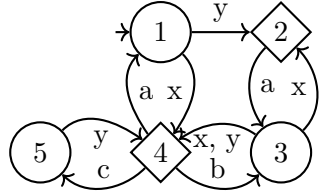
		
		\begin{figure} 
			\centering
		\begin{tikzpicture}[every node/.style={inner sep=2pt}, font=\small]
			\node (piLabel) [anchor=base west] at (0,0) {$\pi=$};
			
			\matrix (m) [matrix of nodes, anchor=base west, ampersand replacement=\&, row sep=0pt, column sep=-1.2pt] 
			at ($(piLabel.east)+(0.01pt,0)$) {
				$1,$ \& $\{\mathbf{y}\},$ \& $2,$ \& $\{a\},$ \& $3,$ \& $\{\mathbf{x}\},$ \& $2,$ \& $\{a\},$ \& $3,$ \& $\{\mathbf{x},\mathbf{y}\},$ \& $4,$ \& $\{a\},$ \& $1,$ \& $\{\mathbf{y}\},$ \& $2,$ \& $\{a\},$ \& $3,$ \& $\{\mathbf{x}\},$ \& $2,$ \& $\{a\},$ \& $3,$ \& $\{\mathbf{x},\mathbf{y}\},$ \& $\dots$ \\
			};
			
			\draw[thick] ([yshift=-7mm] m-1-1.south west) -- ([yshift=-7mm] m-1-14.south east);
			\draw[thick] ([yshift=-7mm] m-1-1.south west) -- ++(0,2mm);
			\draw[thick] ([yshift=-7mm] m-1-14.south east) -- ++(0,2mm);
			\node[anchor=north] at ($([yshift=-7mm] m-1-1.south west)!0.5!([yshift=-7mm] m-1-14.south east)$)
			{$x$ in at least 2 out of 4 turns of $\E$};
			
			\draw[thick] ([yshift=-14mm] m-1-5.south west) -- ([yshift=-14mm] m-1-18.south east);
			\draw[thick] ([yshift=-14mm] m-1-5.south west) -- ++(0,2mm);
			\draw[thick] ([yshift=-14mm] m-1-18.south east) -- ++(0,2mm);
			\node[anchor=north] at ($([yshift=-14mm] m-1-5.south west)!0.5!([yshift=-14mm] m-1-18.south east)$)
			{$x$ in at least 2 out of 4 turns of $\E$};
			
			\draw[thick] ([yshift=-21mm] m-1-9.south west) -- ([yshift=-21mm] m-1-22.south east);
			\draw[thick] ([yshift=-21mm] m-1-9.south west) -- ++(0,2mm);
			\draw[thick] ([yshift=-21mm] m-1-22.south east) -- ++(0,2mm);
			\node[anchor=north] at ($([yshift=-21mm] m-1-9.south west)!0.5!([yshift=-21mm] m-1-22.south east)$)
			{$x$ in at least 2 out of 4 turns of $\E$};
			
			\draw[thick] ([yshift=7mm] m-1-1.north west) -- ([yshift=7mm] m-1-10.north east);
			\draw[thick] ([yshift=7mm] m-1-1.north west) -- ++(0,-2mm);
			\draw[thick] ([yshift=7mm] m-1-10.north east) -- ++(0,-2mm);
			\node[anchor=south] at ($([yshift=7mm] m-1-1.north west)!0.5!([yshift=7mm] m-1-10.north east)$)
			{$y$ in at most 2 out of 3 turns of $\E$};
			
			\draw[thick] ([yshift=14mm] m-1-5.north west) -- ([yshift=14mm] m-1-14.north east);
			\draw[thick] ([yshift=14mm] m-1-5.north west) -- ++(0,-2mm);
			\draw[thick] ([yshift=14mm] m-1-14.north east) -- ++(0,-2mm);
			\node[anchor=south] at ($([yshift=14mm] m-1-5.north west)!0.5!([yshift=14mm] m-1-14.north east)$)
			{$y$ in at most 2 out of 3 turns of $\E$};
			
			\draw[thick] ([yshift=21mm] m-1-9.north west) -- ([yshift=21mm] m-1-18.north east);
			\draw[thick] ([yshift=21mm] m-1-9.north west) -- ++(0,-2mm);
			\draw[thick] ([yshift=21mm] m-1-18.north east) -- ++(0,-2mm);
			\node[anchor=south] at ($([yshift=21mm] m-1-9.north west)!0.5!([yshift=21mm] m-1-18.north east)$)
			{$y$ in at most 2 out of 3 turns of $\E$};
			
			\draw[thick] ([yshift=28mm] m-1-13.north west) -- ([yshift=28mm] m-1-22.north east);
			\draw[thick] ([yshift=28mm] m-1-13.north west) -- ++(0,-2mm);
			\draw[thick] ([yshift=28mm] m-1-22.north east) -- ++(0,-2mm);
			\node[anchor=south] at ($([yshift=28mm] m-1-13.north west)!0.5!([yshift=28mm] m-1-22.north east)$)
			{$y$ in at most 2 out of 3 turns of $\E$};
		\end{tikzpicture}
		\caption{Illustration of the \enquote{sliding window}-property of window counting constraints: Each constraint needs to be fulfilled at each part of a play. Actions in $\pi$ played by $\E$ are highlighted in bold. $\pi$ is a prefix of a play in the game graph shown in Figure~\ref{fig:example_countingConstraints}.}
		\label{fig:prefix_fulfilling_countstraint}
		\end{figure}
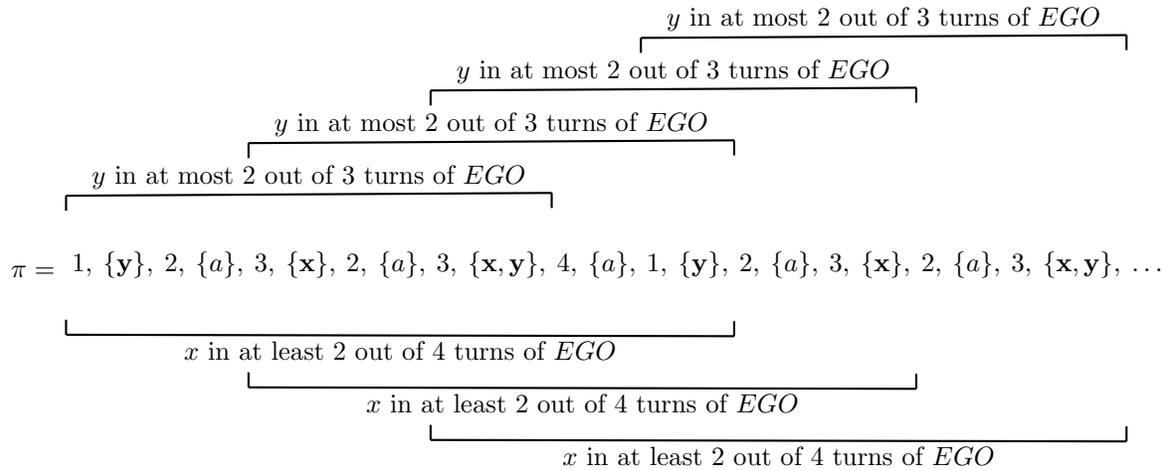
		
		In a (zero-sum) game with counting constraints, the $\E$ player needs to satisfy all of its counting constraints in order to win the game. 
		We assume that the player $\A$  also plays according to its counting constraints. This is an assumption on the rationality of $\A$: $\A$ still plays antagonistic against $\E$, but not if that means to violate the own specifications. For now, we only consider games in which $\A$ can always find a way to satisfy its counting constraints. In particular, $\E$ cannot force $\A$ into a constraint violation.
		
		\begin{defi}[Games with Counting Constraints]\label{def:gameWithCounstraints}
			A two-player \textbf{game with counting constraints} is defined as  $G = (A, \text{Win}, CC)$, where 
			\begin{itemize}
				\item $A = (S, s_0, S_{\E}, S_{\A}, \Sigma_{\E}, \Sigma_{\A}, \rightarrow)$ is a two-player finite-state game graph.
				\item \text{Win} is a winning condition for $\E$ on the game graph $A$.
				\item $CC$ is a finite set of counting constraints.
				\item $\A$ cannot be forced into constraint violations: For each prefix $\pi(n) = \pi_0 \sigma_0 \pi_1 \dots \pi_{n-1}$\linebreak $\sigma_{n-1} \pi_n$, $n \in 2\N+1$, of a play on $G$ that satisfies all counting constraints of $\A$, there exists $(\pi_n, a , \pi_{n+1})\in \rightarrow$, such that $\pi_0 \sigma_0 \pi_1 \dots \pi_{n-1} \sigma_{n-1} \pi_n a \pi_{n+1}$ is also a prefix of a play on $G$ that satisfies the counting constraints of $\A$. 
			\end{itemize}
			Player $\E$ wins a play on $G$, if $\A$ violates one of its constraints in the play or if Win is fulfilled and $\E$ satisfies all of its counting constraints on the play.
			A strategy $\varphi$ of $\E$ is winning for $\E$ (or a \textbf{winning strategy} of $\E$), if $\E$ wins all plays in $O(G, \varphi)$.
		\end{defi}
		
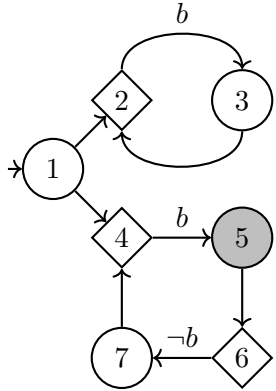
\begin{figure} 
			\centering
			\begin{tikzpicture}[node distance={16mm}, thick, scale=.6] 
				\node[ego] (1) {$1$}; 
				\node[alter] (2) [above right=1, node distance=2.5cm] {$2$}; 
				\node[ego] (3) [right of=2] {$3$};
				\node[alter] (4) [below right=1, node distance=2.5cm] {$4$};
				\node[ego] (5) [right of=4, fill=lightgray] {$5$};
				\node[alter] (6) [below of=5] {$6$};
				\node[ego] (7) [below of=4] {$7$};
				
				\draw[->] (-1,0) -- (1);  
				\draw[->] (1) to node[midway, right] {} (2); 
				\draw[->] (2) to [out=90,in=90]  node[midway, above] {$b$} (3); 
				\draw[->] (3) to [out=270,in=270]  node[midway, below] {} (2);  
				\draw[->] (1) to node[midway, above] {} (4); 
				\draw[->] (4) to [out=0,in=180]  node[midway, above] {$b$} (5); 
				\draw[->] (5) to [out=270,in=90]  node[midway, below] {} (6); 
				\draw[->] (6) to [out=180,in=0]  node[midway, above] {$\neg b$} (7); 
				\draw[->] (7) to [out=90,in=270]  node[midway, left] {} (4);
			\end{tikzpicture} 
			\caption{Graph for a safety game that cannot be used in a game with the counting constraint \enquote{$\A$ plays $b$ at least 1 time in 1 turn.}. Circles represent states controlled by $\E$, diamond-shaped states are controlled by $\A$. The state colored in gray marks an unsafe state. When the lower subgraph is entered, $\A$ cannot fulfill its counting constraint, violating the fourth property in Definition~\ref{def:gameWithCounstraints}.}
			\label{fig:invalid_game}
		\end{figure}

		The fourth property of a game with counting constraints seems counter-intuitive, since it restricts the considered games to those in which $\A$ can basically ensure in each step to satisfy all of its constraints without any need of planning ahead. The main reason for introducing this property is as follows:
		We are not interested in plays that are only won by $\E$ due to forcing $\A$ to lose due to violated counting constraints.  Dismissing the property would lead to situations like depicted in Figure~\ref{fig:invalid_game}: The $\E$-player does not have any counting constraints to obey but needs to avoid an unsafe state (colored in gray). $\A$ has the counting constraint \enquote{$\A$ plays $b$ at least 1 time in 1 turn.} (hence, $\A$ needs to play $b$ in each of its turns). In the initial state, choosing to enter the upper sub-graph lets $\E$ win. However, taking the transitions to the lower subgraph and consequently visiting an unsafe state would also lead to a win for $\E$, since $\A$ is violating its constraint in the resulting play.
		This problem of winning by falsifying the assumptions, but not leading to intended behavior of the player is well-known \cite{MajumdarPS19}, \cite{BloemEJK14}. The fourth property of Definition~\ref{def:gameWithCounstraints} is our way to exclude such trivial solutions inherent to the problem space of synthesis.
		The setting fits especially for applications in which the behavior of $\E$ is influenced by the behavior of the environment (represented by $\A$), but not the other way around. An example for that is a transport robot on a factory floor that is supposed to transport materials to and from machines. Those machines usually have a buffer for materials, such that their behavior is independent of the robot (assuming that an empty input buffer or overfull output buffer would be preceded by a violation of the robots specification). However, which transport tasks need to be done by the robot highly depends on the machines.
		
		In Section~\ref{sec:discussion}, we sketch on which other synthesis settings than the zero-sum-setting is planned as future work as alternative of this restrictive property in the game definition.
		
		\begin{exa} \label{exa:gameWithCountingConstraints}
			The game graph shown in Figure~\ref{fig:example_countingConstraints} together with the counting constraints as given in Example~\ref{exa:countingConstraints} and safety winning condition with all states being safe is a game with counting constraints. Due to its counting constraint, $\A$ will not go from state 4 to state 3 more than once in a row. Hence, $\E$ can ensure to satisfy its counting constraints, e.g. with the following winning strategy: In state 1, play $y$ if $\E$ did not already play $y$ in the last two turns (which would be the case if state 5 was reached before), otherwise play $x$. In state 3, play $x$, if the last own turn was $y$. Otherwise, play $\{x, y\}$.
		\end{exa}
		
		The following lemma characterizes the size of a game graph encoding the combination of a (smaller) graph and a set of counting constraints. 
		\begin{lem}\label{lem:expBlowupOfSitGraph}
			Let  $G = (A, \text{Win}, CC)$ be a game with counting constraints. Let $A'$ be the expanded game graph that encodes the combination of the game graph $A$ with the set $CC$ of counting constraints. The number of states of the game graph $A$ is exponential in the sizes of the counting constraints in $CC$, where the size of a constraint $CC_{m}(p, a, k, l)$, $m \in \{min, max\}$, is the magnitude of the parameter $l$.
		\end{lem}
		\begin{proof}
			For a constraint of the form $CC_{min}(p, a, k, l) \in CC$, each state $s$ in $A$ that is controlled by player $p$ is enriched with information on when in the last $l$ turns the action $a$ was played. Assume this information is encoded as binary sequence of length $l$, in which a \enquote{1} in the $i$th position expressing that action $a$ was played $i$ turns ago ($1\leq i \leq l$). There exist $\sum_{n=k}^{l} \binom{l}{n}$ different sequences of the length $l$ that represent the constraint fulfillment for the last $l$ turns. Thus each state $s$ of the original game graph $A$ is refined into $\sum_{n=k}^{l} \binom{l}{n}$ states in the combined game graph $A'$ (plus some more states for encoding the beginning of the game, in which there are not yet $l$ turns played). For a counting constraint of the form $CC_{max}(p, a, k, l)$, a state in $A$ is refined into $\sum_{n=0}^{k} \binom{l}{n}$ states in $A'$. For a set of constraints, the number of states in $A'$ that refine a state in $A$ is the product of the individual sums of each constraint and the number of states in the original graph. Hence, the number of states in $A'$ is roughly exponential in the size of the product of all constraint lengths, leading to exponential costs for strategy synthesis on $A'$. 
		\end{proof}
		Note that for some games, a memory-efficient winning strategy for the player $\E$ exists. For example, a winning strategy for a game with trivial game graph graph $A$ in which $\E$ can play one of its actions in each turn without further restriction combined with the constraints $CC_{min}(\E, a, k, l)$ and $CC_{min}(\E, b, l-k, l)$ can be realized by first playing $a$ in $k$ turns, followed by playing $b$ in the next $l-k$ turns, requiring only one counter over $l$ values. The combination of counting constraints with a non-trivial game graph raises the need for more complex strategies. 
		Forward-synthesis algorithms aim to avoid the full expansion of the graph, heuristically leading to savings in the synthesis costs. The following section explains how the structure of counting constraints can contribute to this goal. 
		\section{Incremental Synthesis with Counting Constraints}\label{sec:iteratedSynthesis}
		
		The key advantage of counting constraints for synthesis is their monotonicity property: If $\E$ has a strategy, such that $\E$ plays an action $a$ at most $k$ times in $l$ turns (i.e.\ the strategy satisfies $CC_{max}(\E, a, k, l)$), then $\E$ also plays $a$ at most $k$ times in $l-1$ turns (i.e. the strategy satisfies $CC_{max}(\E, a, k, l-1)$).  
		In other words: The existence of a winning strategy for a game with counting constraint $CC_{max}(\E, a, k, l-1)$ is a necessary condition for the winning strategy for a game with $CC_{max}(\E, a, k, l)$. Moreover, only a strategy that fulfills $CC_{max}(\E, a, k, l-1)$ can also fulfill $CC_{max}(\E, a, k, l)$.
		From an algorithmic perspective, it is more favorable to search for strategies that satisfy $CC_{max}(\E, a, k, l-1)$ then for strategies that satisfy $CC_{max}(\E, a, k, l)$, since the graph that encodes the first (shorter) constraint is smaller than the one that encodes the latter (longer) constraint. Intuitively, this is caused by more memory that is needed for remembering the last $l$ own turns instead of only $l-1$ turns. The synthesis idea is related to the incremental approach used by synthesis with antichains \cite{FiliotJR09}.\\
		For a counting constraint of the form  $CC_{min}(\E, a, k, l-1)$ (\enquote{$\E$ plays action $a$ at least $k$ times out of $l-1$ of its turns}), we can conclude that if a strategy fulfills the constraint, it automatically also fulfills the larger constraint  $CC_{min}(\E, a, k, l)$. Hence, if we already have a strategy that fulfills $CC_{min}(\E, a, k, l-1)$, there is no need to search for a strategy that fulfills $CC_{min}(\E, a, k, l)$, which is more computationally demanding.  
		
		\begin{thm}\label{theorem:monotony}
			Let $G = (A, \text{Win}, CC)$ be a two-player game with counting constraints. 
			\begin{enumerate}
				\item \label{theorem:monotony:validity} $(A, \text{Win}, CC')$ is also a two-player game with counting constraints for $CC'$, if $CC$ and $CC'$ only differ in constraints on $\E$.
				\item \label{theorem:negative_monotone}For $CC_{max}(\E, a, k, l) \in CC$ holds: If $\varphi$ is a winning strategy for $\E$ on $G$, then it is also a winning strategy for $\E$ on $G'$, where $G'$ equals $G$ except that $CC_{max}(\E, a, k, l)$ is exchanged by $CC_{max}(\E, a, k, l-1)$.
				\item \label{theorem:positive_monotone}For $CC_{min}(EGO, a, k, l) \in CC$ holds: If $\varphi$ is a winning strategy for $\E$ on $G$, then it is also a winning strategy for $\E$ on $G'$, where $G'$ equals $G$ except that $CC_{max}(\E, a, k, l)$ is exchanged by $CC_{max}(\E, a, k, l+1)$.
				\qed
			\end{enumerate}
		\end{thm}
		
		Since the proof is straightforward, we omit it here.
		It is also possible to vary the $k$ parameter in the constraints instead of $l$ with similar conclusions. 
		
		Theorem~\ref{theorem:monotony} provides the backbone of the presented synthesis approach. However, parts \ref{theorem:negative_monotone} and \ref{theorem:positive_monotone} of the theorem are not directly applicable in the same synthesis procedure, since their monotonicity works in different directions. Incrementing over constraints of the form $CC_{min}(\E, a, k, l)$ identifies which parts of a game are already winnable with smaller constraints and will later be used to prune a constructed graph to those states that are not already known to be in the winning region of the game. Incrementing over constraints of the form $CC_{max}(\E, a, k, l)$ identifies which parts of a game will never become winnable by increasing the constraint length and can be used to prune a constructed graph to those states that are not already known to be outside the winning region of the game. 
		In the scope of this article, we suggest to overcome this problem by translating constraints of the form $CC_{max}(\E, a, k, l)$ to constraints of the other form.
		
		\begin{lem} \label{lemma:translatonOfConstraints}
			Let $G = (A, \text{Win}, CC)$ be a two-player game with counting constraints. For each (infinite) play on $G$ hold the following implications:
					
			\begin{tikzcd}[column sep=1.5cm, row sep=1.0cm, cells={nodes={align=center}}]
				\begin{tabular}[c]{c}
					The player $\E$ plays $a$\\
					at most $k$ times\\
					out of $l$ of its own turns.
				\end{tabular} 
				\arrow[r, shorten <=0.5em, shorten >=0.5em, Rightarrow] 
				\arrow[d, shorten <=0.5em, shorten >=0.5em, Leftrightarrow]
				& 
				\begin{tabular}[c]{c}
					The player $\E$ plays $a$\\
					at most $k+1$ times\\
					out of $l+1$ of its own turns.
				\end{tabular} 
				\arrow[d, shorten <=0.5em, shorten >=0.5em, Leftrightarrow] \\
				\begin{tabular}[c]{c}
					The player $\E$ plays $\neg a$\\
					at least $l-k$ times\\
					out of $l$ of its own turns.
				\end{tabular} 
				\arrow[r, shorten <=0.5em, shorten >=0.5em, Rightarrow]
				& 
				\begin{tabular}[c]{c}
					The player $\E$ plays $\neg a$\\
					at least $l-k$ times\\
					out of $l+1$ of its own turns.
				\end{tabular}
			\end{tikzcd}
		\end{lem}
		Combining Theorem~\ref{theorem:monotony} and Lemma~\ref{lemma:translatonOfConstraints} yields for later increments:
		Instead of directly considering a counting constraint of the form 
		\enquote{The player $\E$ plays $a$ at most $k$ times out of $l$ of its own turns.},
		we can start with  
		\enquote{The player $\E$ plays $\neg a$ at least $l-k$ times out of $l-k$ of its own turns.}, 
		incrementally increase the constraint length to 
		\enquote{The player $\E$ plays $\neg a$ at least $l-k$ times out of $l$ of its own turns.},
		which translates back to the original counting constraint.
		
		With each increment, the number of previously made turns that need to be memorized is increasing. We introduce situation graphs as a mean to encode memory into game graphs, such that a winning strategy on a situation graph is a positional strategy, which is generally not the case for winning strategies for games with window counting constraints. 
		Instead of encoding the whole history of a play up to a given memory size, we focus on memorizing exactly the relevant information for counting constraints. We call a state in the situation graph a situation. In a nutshell, a situation is a state of the game graph $A$ combined with the counting constraints-relevant part of the history on how the state was reached. 
		
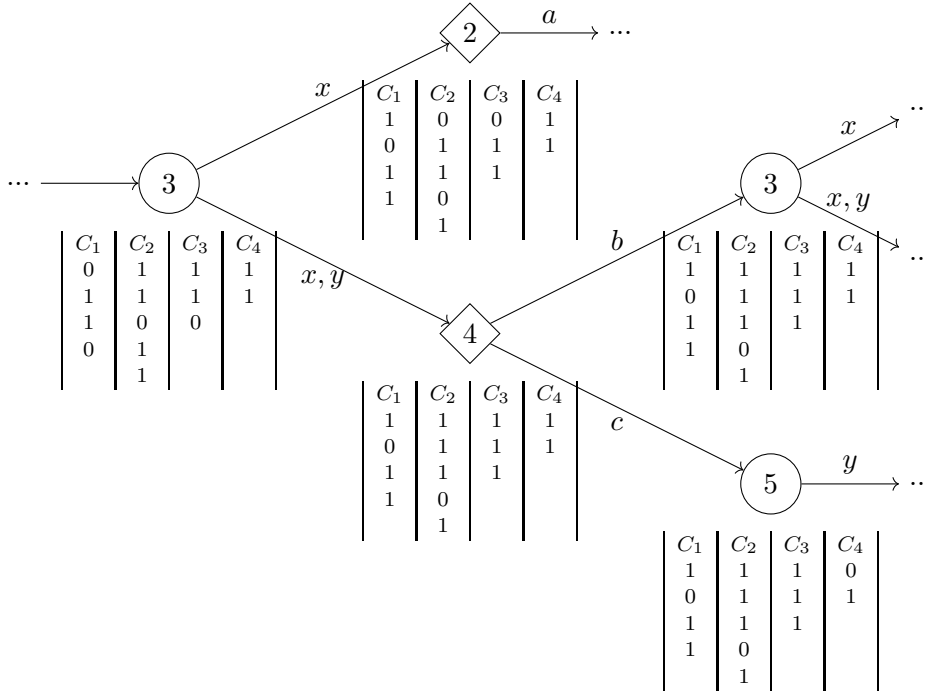
\begin{figure}
		\begin{tikzpicture}[node distance=1cm, auto]
			
			\node[ego] (node1) at (0,0) {3};
			\node[alter] (node2) at (4,2) {2};
			\node[alter] (node3) at (4,-2) {4};
			\node[ego] (node4) at (8,0) {3};
			\node[ego] (node5) at (8,-4) {5};
			\node (dummy2) at (6,2) {...};
			\node (dummy41) at (10,1) {...};
			\node (dummy42) at (10, -1) {...};
			\node (dummy5) at (10, -4) {...};
			
			\node (source) at (-2,0) {...};
			\draw[->] (source) -- (node1);
			
			\draw[->] (node1) -- node[above] {$x$} (node2);
			\draw[->] (node1) -- node[below] {$x, y$} (node3);
			
			\draw[->] (node3) -- node[above] {$b$} (node4);
			\draw[->] (node3) -- node[below] {$c$} (node5);
			
			\draw[->] (node2) -- node[above] {$a$} (dummy2);
			\draw[->] (node4) -- node[above] {$x$} (dummy41);
			\draw[->] (node4) -- node[above] {$x, y$} (dummy42);
			\draw[->] (node5) -- node[above] {$y$} (dummy5);
			
			\node[below=0.1cm of node1] (table1) {\scriptsize %
				\begin{tabular}{|c|c|c|c|}
					$C_1$ & $C_2$ & $C_3$ & $C_4$ \\ 
					0     & 1     & 1     & 1     \\ 
					1     & 1     & 1     & 1     \\ 
					1     & 0     & 0     &       \\ 
					0     & 1     &       &       \\ 
						  & 1     &       &       \\
				\end{tabular}
			};
			
			\node[below=0.1cm of node2] (table2) {\scriptsize
				\begin{tabular}{|c|c|c|c|}
					$C_1$ & $C_2$ & $C_3$ & $C_4$ \\ 
					1     & 0     & 0     & 1     \\ 
					0     & 1     & 1     & 1     \\ 
					1     & 1     & 1     &       \\ 
					1     & 0     &       &       \\ 
						  & 1     &       &       \\ 
				\end{tabular}
			};
			
			\node[below=0.1cm of node3] (table3) {\scriptsize
				\begin{tabular}{|c|c|c|c|}
					$C_1$ & $C_2$ & $C_3$ & $C_4$ \\ 
					1     & 1     & 1     & 1     \\ 
					0     & 1     & 1     & 1     \\ 
					1     & 1     & 1     &       \\ 
					1     & 0     &       &       \\ 
						  & 1     &       &       \\ 
				\end{tabular}
			};
			
			\node[below=0.1cm of node4] (table4) {\scriptsize
				\begin{tabular}{|c|c|c|c|}
					$C_1$ & $C_2$ & $C_3$ & $C_4$ \\ 
					1     & 1     & 1     & 1     \\ 
					0     & 1     & 1     & 1     \\ 
					1     & 1     & 1     &       \\ 
					1     & 0     &       &       \\ 
						  & 1     &       &       \\ 
				\end{tabular}
			};
			
			\node[below=0.1cm of node5] (table5) {\scriptsize
				\begin{tabular}{|c|c|c|c|}
					$C_1$ & $C_2$ & $C_3$ & $C_4$ \\ 
					1     & 1     & 1     & 0     \\ 
					0     & 1     & 1     & 1     \\ 
					1     & 1     & 1     &       \\ 
					1     & 0     &       &       \\ 
						  & 1     &       &       \\ 
				\end{tabular}
			};
			
		\end{tikzpicture}
		\caption{Snippet of a situation graph for the game from Example~\ref{exa:gameWithCountingConstraints}. A state (\enquote{situation}) of the game graph is shown as a combination of a state from the original game graph (Figure~\ref{fig:example_countingConstraints}) with a history that allows to check counting constraint fulfillment in the situation. For better readability, the history is written as table with the name of the counting constraints as header: $C_1 =$ \enquote{The player $\E$ plays $x$ at least $2$ times out of $4$ of its own turns.}, $C_2 = $ \enquote{The player $\E$ plays $y$ at least $1$ times out of $5$ of its own turns.}, $C_3 = $ \enquote{The player $\E$ plays $y$ at most $2$ times out of $3$ of its own turns.}, $C_4 = $ \enquote{The player $\A$ plays $a \vee b$ at least $1$ time out of $2$ of its own turns.}.}
		\label{fig:situationgraphExample}
		\end{figure}

		\begin{exa} \label{exa:situationgraphExample}
			Figure~\ref{fig:situationgraphExample} shows a small part of the situation graph for the game introduced in Example~\ref{exa:countingConstraints} and Figure \ref{fig:example_countingConstraints}. A state from the game graph in Figure~\ref{fig:example_countingConstraints} combined with a history (presented as table in Figure~\ref{fig:situationgraphExample}) is a situation of the situation graph. The left-most state of Figure~\ref{fig:situationgraphExample} can be interpreted as follows: When this state is reached in the situation graph, a corresponding play in the original game is in state 3. In $\E$s last turn, the player did not satisfy the action in counting constraint $C_1 =$ \enquote{The player $\E$ plays $x$ at least $2$ times out of $4$ of its own turns.} (\enquote{0} as first entry in the $C_1$-column), but satisfied the actions in $C_2$ and $C_3$ (\enquote{1} as first entry in the $C_2$- and $C_3$-columns). In the second to last turn of $\E$, all of the actions in the $\E$-constraints were fulfilled (\enquote{1} as second entry in the $C_1$-, $C_2$- and $C_3$-columns). For each counting constraint, only the amount of last turns relevant for the constraint are considered in the situation graph. Hence, some of the entries in the tables of Figure~\ref{fig:situationgraphExample} are left empty. It can be seen that the leftmost and the uppermost situation fulfill all counting constraints. All other situations violate the constraint  $C_3 = $ \enquote{The player $\E$ plays $y$ at most $2$ times out of $3$ of its own turns.}, since $\E$ played $y$ in all of its last three turns (three times \enquote{1}-entries in the $C_3$-columns). Hence, no winning strategy of $\E$ will suggest to play $x, y$ in state 3 of the game in Figure~\ref{fig:example_countingConstraints} when the state was reached as induced by the history in the leftmost situation of Figure~\ref{fig:situationgraphExample}.
		\end{exa}

		\begin{defi}[Situation Graph]\label{def:situationGraph}
			Let $G = (A, \text{Win}, CC)$ a game with counting constraints with $A = (S, s_0, S_{\E}, S_{\A}, \Sigma_{\E}, \Sigma_{\A}, \rightarrow)$. To ease the following notations, fix some order in the counting constraints $CC = \{CC_1, \dots, CC_n\}$. \\
			For each counting constraint $C:=CC_m(p, a, k, l)$ with $m\in \{min, max\}$, player $p\in \{\E, \A\}$, $a$ being a logical formula over $\Sigma_p$ as defined in Definition~\ref{def:windowCountingConstraints} and $k, l \in \N$ with $k\leq l$, define a mapping 
			\[
				h_C\colon \{0,1,\text{none}\}^l \times \mathcal{P}(\Sigma_p) \to \{0,1,\text{none}\}^l, \, 
				((v_1,\dots,v_l), act) \mapsto \biggl\{
				\begin{aligned}
					(1,v_1,\dots,v_{l-1}), &\text{ if } act \models a\\
					(0,v_1,\dots,v_{l-1}), &\text{ else.}
				\end{aligned} 
			\]
			This mapping serves for encoding how constraint fulfillment evolves when a player does a turn in a play.

			A situation is a tuple $(s, H)$ with $s\in S$ being a state in $A$, $H \in \prod_{i=1}^n \text{codom}(h_{C_i})$ and $\text{codom}(f)=Y$ denoting the codomain of a function $f\colon X\rightarrow Y$.  
			Denote the set of all situations by $\tilde{S}$, $\tilde{S}_{\E}$ the set of situations with state (in $G$) controlled by $\E$ and $\tilde{S}_{\A}$ the set of situations with a state controlled by $\A$.
			Define a mapping 
			\begin{align*}
				\hookrightarrow' \colon \Big( \tilde{S}_{\E} \times \mathcal{P}(\Sigma_{\E}) \Big) \cup \Big( \tilde{S}_{\A} \times \mathcal{P}(\Sigma_{\A}) \Big) &\to \tilde{S} \\ ((s, (v_1,\dots,v_q)),act) & \mapsto (s', (w_1, \dots, w_n))
			\end{align*}
			with $(s,act,s')\in \rightarrow$ and
			\[
				w_i = \biggl\{
					\begin{aligned}
						h_{CC_i}(v_i, act), &\text{ if } s \text{ is controlled by the player that has to satisfy } CC_i\\
						v_i, &\text{  else.}
					\end{aligned} 
			\]
			The transition $\hookrightarrow'$ defines how to get from one situation to another when using the transition $\rightarrow$ in $G$.\\
			A situation $(s, (v_1, \dots, v_n))$ satisfies a counting constraint $CC_i =  CC_m(p, a, k, l)$, if for the history-part $v_i = (v'_1, \dots, v'_l) \in \{0,1,\text{none}\}^l $ holds:
			\[ | \{ j \in \{1, \dots, l\} \, | \, v'_j \neq 0 \} | \quad \biggl\{
			\begin{aligned}
				\geq k, &\text{ if } m=min\\
				\leq k, &\text{ if } m=max.
			\end{aligned}
			\]
			
			The \textbf{situation graph} of $G$ is the two-player finite game graph 
			\[Sit = (S', s_{init},S'_{\E}, S'_{\A}, \Sigma_{\E}, \Sigma_{\A}, \hookrightarrow, \text{Win}', CC)\] 
			with
			\begin{itemize}
				\item initial state being the situation $s_{init}=(s_0, H_{init})$ with all entries in $H_{init}$ being $none$,
				\item transition relation $\hookrightarrow \subseteq \tilde{S} \times (\mathcal{P}(\Sigma_{\E}) \cup \mathcal{P}(\Sigma_{\A})) \times \tilde{S}$ with $((s, H),act,(s', H'))\in \hookrightarrow$ if and only if $(s, act, s')\in \rightarrow$ and $\hookrightarrow'((s,H), act)=(s',H')$,
				\item set of states $S'$ being all situations that are reachable from $s_{init}$ via $\hookrightarrow$ and in which all counting constraints on $\A$ are satisfied.
				\item $S'_{p}\subseteq S'$ the states $(s,H)$ that are controlled by player $p\in \{\E,\A\}$, that is $s\in S_p$.
				\item $\text{Win}'$ the winning condition for $\E$: Extend the winning condition $\text{Win} \subseteq S^{\omega}$ in $G$ to the situation graph by defining $\text{Win}' = \{ (s_i,H_i)_{i \in \N} \, |\, (s_i)_{i \in \N} \in \text{Win} \text{ and for each } i \in \N \text { holds: } (s_i,H_i) \text{ satisfies all counting constraints from $CC$ on } \E \}$
			\end{itemize}
			The winning region of $\E$ in the situation graph is the set of states $\tilde{S}\subseteq S'$ from which $\E$ has a winning strategy.
		\end{defi}
		
		The incremental synthesis approach avoids to construct the full situation graph, which is exponential in the constraint lengths (Lemma~\ref{lem:expBlowupOfSitGraph}, as the expanded game graph in the lemma is basically a situation graph without details on the notation). In fact, we will later use a slightly modified version of the situation graph that allows us to actually apply any synthesis algorithm that is also applicable on the original game $G$ without counting constraints.

\begin{figure}
			\begin{tikzpicture}
				\node[ego] (node1) at (5,0) {3};
				\node[below=0.1cm of node1] (table1) {\scriptsize %
					\begin{tabular}{|c|c|c|c|}
						$C'_1$ & $C'_2$ & $C_3$ & $C_4$ \\ 
						0     & 1     & 1     & 1     \\ 
						1     & 1     & 1     & 1     \\ 
						1     & 0     & 0     &       \\ 
						0     & 1     &       &       \\ 
							  & 1     &       &       \\
					\end{tabular}
				};
				\node[below=0.1cm of table1] (constr1) {\scriptsize %
					\begin{tabular}{c}
						$C'_1 = CC_{min}(\E, x, 2, 4)$ \\
						$C'_2 = CC_{min}(\E, y, 1, 5)$ \\
						$C_3 = CC_{max}(\E, y, 2, 3)$ \\
						$C_4 = CC_{min}(\A, a \lor b, 1, 2)$ \\
					\end{tabular}
				};
				
				\node[ego] (node2) at (10,0) {3};
				\node[below=0.1cm of node2] (table2) {\scriptsize %
					\begin{tabular}{|c|c|c|c|}
						$C''_1$ & $C''_2$ & $C_3$ & $C_4$ \\ 
						0     & 1     & 1     & 1     \\ 
						1     & 1     & 1     & 1     \\ 
						1     &       & 0     &       \\ 
						 	  &       &       &       \\ 
							  &       &       &       \\
					\end{tabular}
				};
				\node[below=0.1cm of table2] (constr2) {\scriptsize %
					\begin{tabular}{c}
						$C''_1 = CC_{min}(\E, x, 2, 3)$ \\
						$C''_2 = CC_{min}(\E, y, 1, 2)$ \\
						$C_3 = CC_{max}(\E, y, 2, 3)$ \\
						$C_4 = CC_{min}(\A, a \lor b, 1, 2)$ \\
					\end{tabular}
				};
				
				\node[ego] (node3) at (0,0) {3};
				\node[below=0.1cm of node3] (table3) {\scriptsize %
					\begin{tabular}{|c|c|c|c|}
						$C_1$ & $C_2$ & $C_3$ & $C_4$ \\ 
						0     & 1     & 1     & 1     \\ 
						1     & 1     & 1     & 1     \\ 
						1     & 0     & 0     &       \\ 
						0     & 1     &       &       \\ 
						1	  & 1     &       &       \\
					\end{tabular}
				};
				\node[below=0.1cm of table3] (constr3) {\scriptsize %
					\begin{tabular}{c}
						$C_1 = CC_{min}(\E, x, 2, 5)$ \\
						$C_2 = CC_{min}(\E, y, 1, 5)$ \\
						$C_3 = CC_{max}(\E, y, 2, 3)$ \\
						$C_4 = CC_{min}(\A, a \lor b, 1, 2)$ \\
					\end{tabular}
				};
				
				\draw[dotted, ->] (node1) -- node[above] {is extension of} (node2);
				\draw[dotted, ->] (node3) -- node[above] {is extension of} (node1);
			\end{tikzpicture}
			\caption{Examples for situations being extensions of other situations. The notation is the same as in Example~\ref{exa:situationgraphExample} and Figure~\ref{fig:situationgraphExample}.}
			\label{fig:situationExtensionsExample}
			\end{figure}
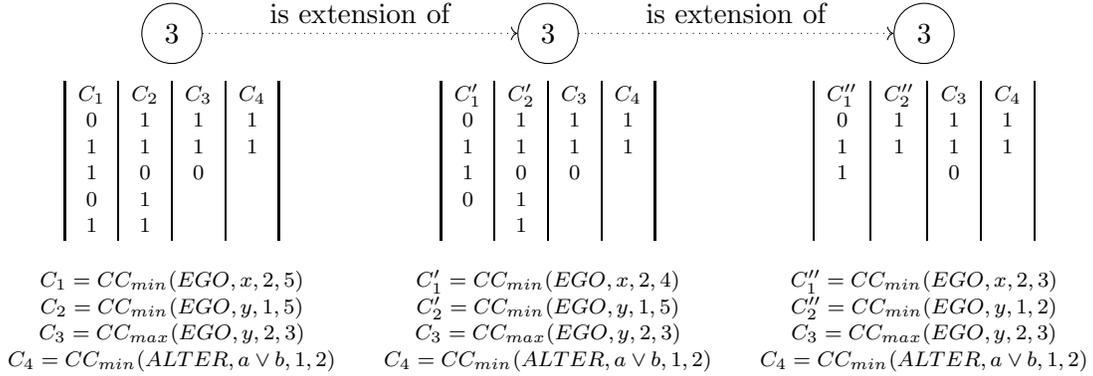

		The general idea is to start with a rather small game by using counting constraints of small lengths and increment over the length. In each increment, a part of the corresponding situation graph is constructed and analyzed and knowledge that can be reused in following increments is identified. This knowledge is determining which parts of the situation graph for the next increment needs to be constructed and which parts can be omitted, relying on \autoref{theorem:monotony}. To this end, we introduce extensions of situations in order to compare situations from one increment with situations in another increment.
		\begin{defi}[Extensions of situations]\label{def:extensionOfSituation}
			Let $(s, H)$ be a situation with counting constraints $CC$, $H=(H_1,\dots, H_n)$, $H_i$ representing the history for counting constraint $CC_i\in CC$. 
			\begin{itemize}
				\item A set $CC'$ of counting constraints is an \textbf{extension of counting constraints} $CC$, if the only difference between $CC$ and $CC'$ are longer counting constraints of the form $CC_{min}(\E, a, k, l)$. Formally, $CC'$ is an extension of $CC$, if $| CC |= | CC' |$ and if for all $CC_i \in CC$, there exists $CC_i'\in CC'$ with
				\[
				 CC_i' = \biggl\{
				\begin{aligned}
					CC_{min}(\E, a, k, l') \text{ with } l' \geq l, &\text{ if } CC_i = CC_{min}(\E, a, k, l) \\
					CC_i, &\text{  else.}
				\end{aligned} 
				\]
				\item A history $H'$ with counting constraints $CC'$ is an \textbf{extension of a history} $H$, if $CC'$ is an extension of $CC$ and $H'$ differs from $H$ only by additional entries for constraints of the form $CC_i = CC_{min}(\E, a, k, l)$.
				Formally: $H'$ is an extension of $H$, if the corresponding counting constraints $CC'$ are an extension of $CC$ and it exists an $X\in (\{0, 1, \text{none}\}^*)^{| CC |}$, such that $H'=H.X$ (element-wise concatenation).
				\item A situation $(s', H')$ is an extension of a situation $(s, H)$, if $s=s'$ and $H'$ is an extension of $H$. In a nutshell, an extension $(s',H')$ of a situation $(s,H)$ provides more information than $(s, H)$ on former turns of $\E$ with respect to constraints of the form $CC_{min}(\E, a, k, l)$ and equals $(s,H)$ in all other aspects.
			\end{itemize}
		\end{defi}
		\begin{exa}
			In Figure~\ref{fig:situationExtensionsExample}, one of the situations of the previous example (Example~\ref{exa:situationgraphExample}, leftmost situation of Figure~\ref{fig:situationgraphExample}) is shown with one extension of the situation and one situation that extends itself.
		\end{exa}
		
		For the scope of this article, we restrict ourselves to increment over constraints of the form $CC_{min}(\E, a, k, l)$ and the usage of Lemma~\ref{lemma:translatonOfConstraints}. 
		The overall synthesis procedure is sketched in Figure~\ref{fig:overview_iteration_min}. We explain the different phases of the incremental synthesis algorithm for a two-player finite-state game with counting constraints $G=(A, \text{Win}, CC)$ based on this figure. 
		In each increment, a synthesis algorithm of the readers choice that can determine the winning region of a game $(A, \text{Win})$, with the chosen winning condition is used. Since the concept of a winning region only makes sense for positionally determined games, we assume that the winning condition \text{Win} is either that of a safety, reachability, Büchi, Co-Büchi or parity game. 
		The winning condition influences how the situation graph is modified in each increment. We checked the modification explicitly for the mentioned winning conditions. However, the approach might be applicable for more types of winning conditions, especially for other positionally determined games. 
		\tikzset{
			decision/.style={diamond, draw, text width=5em, text badly centered, node distance=2.5cm, inner sep=0pt},
			block/.style={rectangle, draw, text width=9em, text centered, rounded corners, minimum height=3em},   
			largeblock/.style={rectangle, draw, text width=18em, text centered, rounded corners, minimum height=3em}, 
			line/.style={draw, -latex'},
			result/.style={rectangle, draw, fill=white!20, text width=8em, text centered, rounded corners, minimum height=3em}
		}
		\begin{figure}
			\centering
			\small
			\centering
			\begin{tikzpicture}[node distance = 0.4cm, auto] 
				\node [largeblock] (translate) {\textbf{Translate} counting constraints of the form $CC_{max}(\E, a, k, l)$ to constraints of the form $CC_{min}(\E, \neg a, l-k, l)$. Denote the resulting set of counting constraints as $\mathscr{C}$.};
				\node [largeblock, below=of translate] (init) {\textbf{Initialize} set $\mathcal{C}$ of counting constraints for the first increment.};
				\node [block, below=of init] (situations) {(Partly) construct \textbf{situation graph}, memory depending on $\mathcal{C}$.};
				\node [block, below=of situations] (adaptsit) {Modify situation graph to \textbf{winning condition} \text{Win}.};
				\node [block, below=of adaptsit] (region) {Determine the \textbf{winning region}.};
				\node [block, left of=region, node distance=3.5cm] (update) {\textbf{Increase} length of counting constraints of the form $CC_{min}(EGO, a, k, l)$ in $\mathcal{C}$.};
				\node [decision, below of=region] (checkinit) {Winning strategy found?};
				\node [decision, below of=checkinit, node distance=2.5cm] (checkconst) {$\mathcal{C} = \mathscr{C}$?};
				\coordinate [below of=checkconst, node distance=2cm] (resultsline) {};
				\node [result, right of=resultsline, node distance=0.5cm] (winning) {Winning strategy found.};
				\node [result, left of=resultsline, node distance=3cm] (losing) {No winning strategy exists.};
				\coordinate [right of=checkinit, node distance=1.5cm] (gonnawin) {};
				\coordinate [below of=checkconst, node distance=1.2cm] (belowcheckconst) {};
				\coordinate [above of=losing, node distance=0.8cm] (gonnalose) {};
				\coordinate [below of=gonnawin, node distance=3.8cm] (abovegonnawin) {};
				\path [line] (translate)  -- (init);
				\path [line] (init) -- (situations); 
				\path [line] (situations) -- (adaptsit);
				\path [line] (adaptsit) -- (region);
				\path [line] (region) -- (checkinit);
				\path [line] (checkinit) -- (checkconst);
				\path [line] (checkconst) -| node [near start] {No} (update);
				\path [line] (update) |- (situations);
				\path [line] (checkinit) -- node {No} (checkconst); 
				\path [line] (checkinit) -|  (gonnawin) -- node [right] {Yes} (abovegonnawin) -| (winning);
				\path [line] (checkconst) -- node {} (belowcheckconst) -- node [above] {Yes} (gonnalose) -| (losing);
			\end{tikzpicture}
			\caption{Overview of the incremental synthesis algorithm for a game with counting constraints $G=(A, \text{Win}, CC)$ with increments over constraints of the form $CC_{min}(\E, a, k, l)$.}
			\label{fig:overview_iteration_min}
		\end{figure}
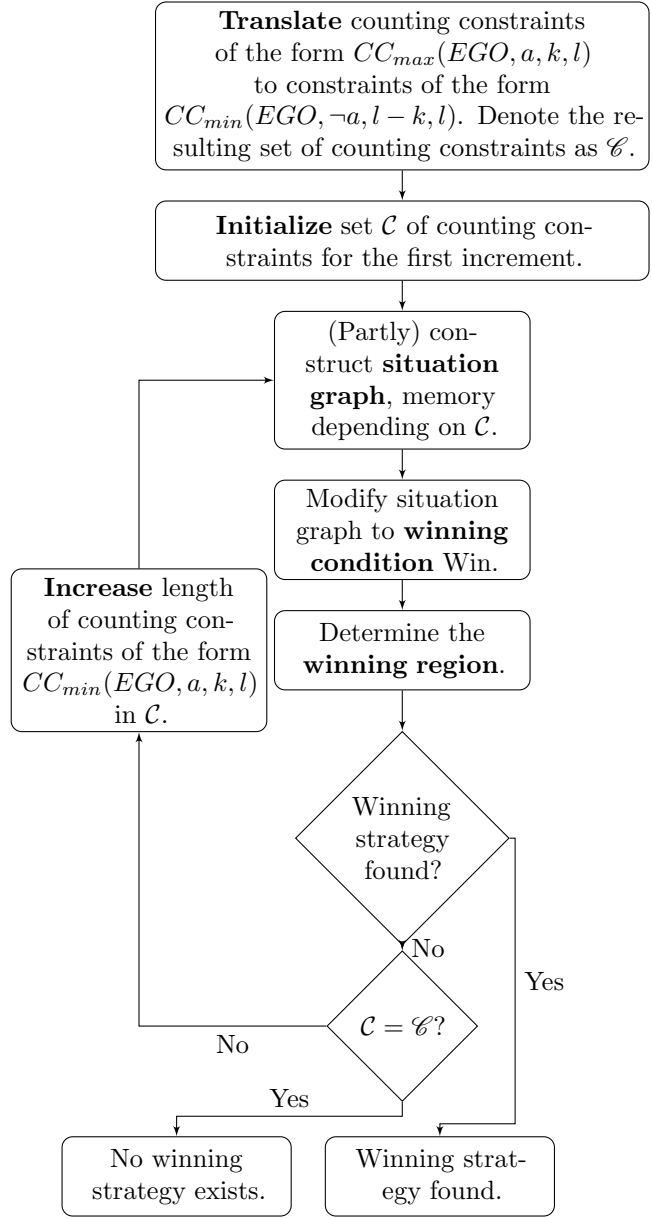
		
		\subsection*{Translate counting constraints}
		(Optionally) Translate window counting constraints in $CC$ of the form $CC_{min}(\E, a, k, l)$ to constraints of the form $CC_{min}(\E, \neg a, l-k, l)$. All other constraints remain as is. Denote the resulting set of counting constraints with $\mathscr{C}$. Lemma~\ref{lemma:translatonOfConstraints} ensures that a winning strategy for $(A, \text{Win}, \mathscr{C})$ is also a winning strategy for $G=(A, \text{Win}, CC)$ and allows to increment over the length of more counting constraints than without translation. Constraints that are not of the form $CC_{min}(\E, a, k, l)$ are considered to be fixed in the synthesis procedure and hence, the required memory in the full length of those constraints is completely encoded into the game. However, this translation process is optional and there are games that are actually solvable quicker without the translation process. This is for example the case when incrementing over a specific constraint does not contribute to extending the winning region for any constraint length other then the full length. 
		\subsection*{Initialize counting constraints for the first increment}
		In this step, it is determined with which combination of constraint lengths the incremental synthesis algorithm shall start. For each constraint in $\mathscr{C}$ of the form $CC_{min}(\E, a, k, l)$, choose a $l' \in \N$ with $k \leq l' \leq l$ and add $CC_{min}(\E, a, k, l')$ to the set $\mathcal{C}$ of starting constraints. All constraints in $\mathscr{C}$ that are not of the form $CC_{min}(\E, a, k, l)$ are added to $\mathcal{C}$ without modification. A canonical implementation for this step is to set all counting constraints to their minimal length, that is: initialize  $CC_{min}(\E, a, k, l')$. Some games might allow for more customized solutions. For instance, if the game graph $A$ shows that $\E$ can only play exactly one atomic action in each of its turns, (that is,  $\rightarrow \subseteq S \times (\Sigma_{\E} \cup \mathcal{P}(\Sigma_{\A})) \times S$ in Definition~\ref{def:twoPlayerGameGraph}), the first increments can be skipped over and  $CC_{min}(\E, a, k, l')$ can be instantiated by $CC_{min}(\E, a, k, l'')$ with $l'' =  \sum_{CC_{min}(\E, \tilde{a}, \tilde{k}, \tilde{l})\in \mathscr{C}} \tilde{k}$. 
		\subsection*{(Partly) construct situation graph for $(A, \text{Win}, \mathcal{C})$}
\begin{figure}
			\centering
			\begin{subfigure}[b]{0.8\textwidth}
				\centering
				\begin{tikzpicture}[node distance=5mm and 12mm, thick, scale=0.8, every node/.style={transform shape}] 
					\node[ddd] (ddd1) {...}; 
					\node[ego] (s1h1) [right=.5cm of  ddd1] {\SH{1}{1}};
					\node[alter] (s2h2) [above right= of s1h1] {\SH{2}{2}};
					\node[ddd] (ddd2a) [above left=2mm and 4mm of s2h2] {...}; 
					\node[ddd] (ddd2) [right=.5cm of s1h1] {...}; 
					\node[alter, EconstViolated] (s3h3) [below right=12mm and 12mm of  s1h1] {\SH{3}{3}};
					\node[ddd] (ddd3) [above right=2mm and 4mm of s3h3] {...}; 
					\node[ddd] (ddd3a) [below left=2mm and 4mm of s3h3] {...}; 
					\node[ego] (s6h6) [right=8mm of s3h3] {\SH{6}{6}};
					\node[ddd] (ddd4a) [above right=2mm and 4mm of s6h6] {...}; 
					\node[ddd] (ddd4) [below left=2mm and 4mm of s6h6] {...}; 
					\node[ego, AconstViolated] (s4h4) [above right= of s2h2] {\SH{4}{4}};
					\node[ddd] (ddd5a) [above left=1mm and 4mm of s4h4] {...}; 
					\node[ddd] (ddd5) [right=.5cm of s2h2] {...}; 
					\node[ego] (s5h5) [below right= of s2h2] {\SH{5}{5}};
					\node[ddd] (ddd6a) [right=.5cm of s5h5] {...}; 
					\node[ddd] (ddd6) [right=.5cm of s4h4] {...}; 
					\node[alter, winnable] (s7h7) [above right=5mm and 20mm of  s5h5] {\SH{7}{7}};
					\node[alter] (s8h8) [below right=5mm and 20mm of  s5h5] {\SH{8}{8}};
					\node[ddd] (ddd7) [above left=2mm and 4mm of s7h7] {...}; 
					\node[ddd] (ddd8) [below right=2mm and 4mm of s7h7] {...}; 
					\node[ddd] (ddd9) [below left=2mm and 4mm of s8h8] {...}; 
					\node[ego] (s9h9) [right= of s7h7] {\SH{9}{9}};
					\node[ddd] (ddd10) [above left=2mm and 4mm of s9h9] {...}; 
					\node[ddd] (ddd11) [right=.5cm of s9h9] {...}; 
					\node[ego] (s10h10) [above right=2mm and 20mm of s8h8] {\SH{10}{10}};
					\node[ddd] (ddd12) [right=.5cm of s8h8] {...}; 
					\node[ddd] (ddd13) [above left=2mm and 4mm of s10h10] {...}; 
					\node[ddd] (ddd14) [right=.5cm of s10h10] {...}; 
					\node[ego, winnable] (s11h11) [below right=2mm and 20mm of s8h8] {\SH{11}{11}};
					\node[ddd] (ddd15) [below left=2mm and 4mm of s11h11] {...}; 
					\node[ddd] (ddd16) [right=.5cm of s11h11] {...}; 
					
					\draw[->] (ddd1) -- (s1h1); 
					\draw[->] (s1h1) -- (s2h2); 
					\draw[->] (ddd2a) -- (s2h2); 
					\draw[->] (s1h1) -- (ddd2);
					\draw[->] (s1h1) -- (s3h3); 
					\draw[->] (s3h3) -- (ddd3); 
					\draw[->] (ddd3a) -- (s3h3); 
					\draw[->] (s3h3) -- (s6h6);
					\draw[->] (ddd4) -- (s6h6); 
					\draw[->] (s6h6) -- (ddd4a); 
					\draw[->] (s2h2) -- (s4h4); 
					\draw[->] (s2h2) -- (ddd5);
					\draw[->] (s2h2) -- (s5h5); 
					\draw[->] (s4h4) -- (ddd6); 
					\draw[->] (s5h5) -- (s7h7);
					\draw[->] (s5h5) -- (s8h8); 
					\draw[->] (s5h5) -- (ddd6a); 
					\draw[->] (ddd7) -- (s7h7); 
					\draw[->] (s7h7) -- (ddd8);
					\draw[->] (s7h7) -- (s9h9); 
					\draw[->] (ddd10) -- (s9h9); 
					\draw[->] (s9h9) -- (ddd11);
					\draw[->] (ddd9) -- (s8h8); 
					\draw[->] (s8h8) -- (s10h10); 
					\draw[->] (s8h8) -- (ddd12);
					\draw[->] (s8h8) -- (s11h11); 
					\draw[->] (ddd13) -- (s10h10); 
					\draw[->] (s10h10) -- (ddd14);
					\draw[->] (ddd15) -- (s11h11); 
					\draw[->] (s11h11) -- (ddd16);
					\draw[->] (ddd5a) -- (s4h4);
					
				\end{tikzpicture} 
				\caption{Snippet of a situation graph conform to Definition~\ref{def:situationGraph}.}
				\label{subfig:sitgraph_nach_def}
			\end{subfigure}
			\begin{subfigure}[b]{0.8\textwidth}
				\centering
				\begin{tikzpicture}[node distance=5mm and 12mm, thick, scale=0.8, every node/.style={transform shape}] 
					\node[ddd] (ddd1) {...}; 
					\node[ego] (s1h1) [right=.5cm of  ddd1] {\SH{1}{1}};
					\node[alter] (s2h2) [above right= of s1h1] {\SH{2}{2}};
					\node[ddd] (ddd2a) [above left=2mm and 4mm of s2h2] {...}; 
					\node[ddd] (ddd2) [right=.5cm of s1h1] {...}; 
					\node[alter, EconstViolated] (s3h3) [below right=12mm and 12mm of  s1h1] {\SH{3}{3}};
					\node[ddd] (ddd3a) [below left=2mm and 4mm of s3h3] {...}; 
					\node[ddd] (ddd5) [right=.5cm of s2h2] {...}; 
					\node[ego] (s5h5) [below right= of s2h2] {\SH{5}{5}};
					\node[ddd] (ddd6a) [right=.5cm of s5h5] {...}; 
					\node[alter, winnable] (s7h7) [above right=5mm and 20mm of  s5h5] {\SH{7}{7}};
					\node[alter] (s8h8) [below right=5mm and 20mm of  s5h5] {\SH{8}{8}};
					\node[ddd] (ddd7) [above left=2mm and 4mm of s7h7] {...}; 
					\node[ddd] (ddd9) [below left=2mm and 4mm of s8h8] {...}; 
					\node[ego] (s10h10) [above right=2mm and 20mm of s8h8] {\SH{10}{10}};
					\node[ddd] (ddd12) [right=.5cm of s8h8] {...}; 
					\node[ddd] (ddd13) [above left=2mm and 4mm of s10h10] {...}; 
					\node[ddd] (ddd14) [right=.5cm of s10h10] {...}; 
					\node[ego, winnable] (s11h11) [below right=2mm and 20mm of s8h8] {\SH{11}{11}};
					\node[ddd] (ddd15) [below left=2mm and 4mm of s11h11] {...}; 
					
					\draw[->] (ddd1) -- (s1h1); 
					\draw[->] (s1h1) -- (s2h2); 
					\draw[->] (ddd2a) -- (s2h2); 
					\draw[->] (s1h1) -- (ddd2);
					\draw[->] (s1h1) -- (s3h3);  
					\draw[->] (ddd3a) -- (s3h3);  
					\draw[->] (s2h2) -- (ddd5);
					\draw[->] (s2h2) -- (s5h5); 
					\draw[->] (s5h5) -- (s7h7);
					\draw[->] (s5h5) -- (s8h8); 
					\draw[->] (s5h5) -- (ddd6a); 
					\draw[->] (ddd7) -- (s7h7); 
					\draw[->] (ddd9) -- (s8h8); 
					\draw[->] (s8h8) -- (s10h10); 
					\draw[->] (s8h8) -- (ddd12);
					\draw[->] (s8h8) -- (s11h11); 
					\draw[->] (ddd13) -- (s10h10); 
					\draw[->] (s10h10) -- (ddd14);
					\draw[->] (ddd15) -- (s11h11); 
					
				\end{tikzpicture} 
				\caption{Snippet of the modified situation graph as constructed in the step \enquote{(partly) construct situation graph}.}
				\label{subfig:sitgraph_shortened}
			\end{subfigure}
			\begin{subfigure}[b]{0.8\textwidth}
				\centering
				\begin{tikzpicture}[node distance=5mm and 12mm, thick, scale=0.8, every node/.style={transform shape}] 
					\node[ddd] (ddd1) {...}; 
					\node[ego] (s1h1) [right=.5cm of  ddd1] {\SH{1}{1}};
					\node[alter] (s2h2) [above right= of s1h1] {\SH{2}{2}};
					\node[ddd] (ddd2a) [above left=2mm and 4mm of s2h2] {...}; 
					\node[ddd] (ddd2) [right=.5cm of s1h1] {...}; 
					\node[alter, EconstViolated] (s3h3) [below right=12mm and 12mm of  s1h1] {\SH{3}{3}};
					\node[ddd] (ddd3a) [below left=2mm and 4mm of s3h3] {...}; 
					\node[ego] (lE) [below right= of s3h3] {$l_E, \boldsymbol{\cdot}\,$};
					\node[alter] (lA) [right= of lE] {$l_A, \boldsymbol{\cdot}\,$};
					\node[ddd] (ddd5) [right=.5cm of s2h2] {...}; 
					\node[ego] (s5h5) [below right= of s2h2] {\SH{5}{5}};
					\node[ddd] (ddd6a) [right=.5cm of s5h5] {...}; 
					\node[alter, winnable] (s7h7) [above right=5mm and 20mm of  s5h5] {\SH{7}{7}};
					\node[ego] (wE) [above right=2mm and 20mm of s7h7] {$w_E, \boldsymbol{\cdot}\,$};
					\node[alter] (wA) [right= of wE] {$w_A, \boldsymbol{\cdot}\,$};
					\node[alter] (s8h8) [below right=5mm and 20mm of  s5h5] {\SH{8}{8}};
					\node[ddd] (ddd7) [above left=2mm and 4mm of s7h7] {...}; 
					\node[ddd] (ddd9) [below left=2mm and 4mm of s8h8] {...}; 
					\node[ego] (s10h10) [above right=2mm and 20mm of s8h8] {\SH{10}{10}};
					\node[ddd] (ddd12) [right=.5cm of s8h8] {...}; 
					\node[ddd] (ddd13) [above left=2mm and 4mm of s10h10] {...}; 
					\node[ddd] (ddd14) [right=.5cm of s10h10] {...}; 
					\node[ego, winnable] (s11h11) [below right=2mm and 20mm of s8h8] {\SH{11}{11}};
					\node[ddd] (ddd15) [below left=2mm and 4mm of s11h11] {...}; 
					
					\draw[->] (ddd1) -- (s1h1); 
					\draw[->] (s1h1) -- (s2h2); 
					\draw[->] (ddd2a) -- (s2h2); 
					\draw[->] (s1h1) -- (ddd2);
					\draw[->] (s1h1) -- (s3h3); 
					\draw[->] (s3h3) -- (lE); 
					\draw[->] (lE) to [out=45,in=135] (lA); 
					\draw[->] (lA) to [out=225,in=315] (lE); 
					\draw[->] (ddd3a) -- (s3h3); 
					\draw[->] (s2h2) -- (ddd5);
					\draw[->] (s2h2) -- (s5h5);  
					\draw[->] (s5h5) -- (s7h7);
					\draw[->] (s5h5) -- (s8h8); 
					\draw[->] (s5h5) -- (ddd6a); 
					\draw[->] (ddd7) -- (s7h7); 
					\draw[->] (s7h7) -- (wE); 
					\draw[->] (wE) to [out=45,in=135] (wA); 
					\draw[->] (wA) to [out=225,in=315] (wE); 
					\draw[->] (ddd9) -- (s8h8); 
					\draw[->] (s8h8) -- (s10h10); 
					\draw[->] (s8h8) -- (ddd12);
					\draw[->] (s8h8) -- (s11h11); 
					\draw[->] (ddd13) -- (s10h10); 
					\draw[->] (s10h10) -- (ddd14);
					\draw[->] (ddd15) -- (s11h11); 
					\draw[->] (s11h11) to [out=0,in=-45] (wA); 
				\end{tikzpicture} 
				\caption{Snippet of the situation graph with winning and losing sinks incorporated as described in step \enquote{modify situation graph to winning condition}.}
				\label{subfig:sitgraph_with_sinks}
			\end{subfigure}
		\caption{Illustration on how the situation graph is modified in the incremental synthesis over counting constraint length as depicted in Figure~\ref{fig:overview_iteration_min}. Circles represent situations controlled by $\E$, diamond-shaped situations are controlled by $\A$. Situations colored in gray represent extensions of situations in the winning region of a previous increment. Situations marked with doubled lines are considered to have a history that violates a constraint of $\E$. Similarly, histories of situations with dashed lines are violating a constraint of $\A$.}
		\label{fig:fromSitGraphToSinks}
		\end{figure}
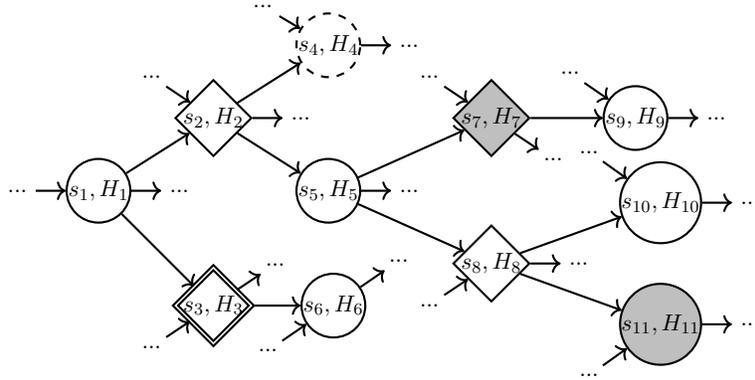
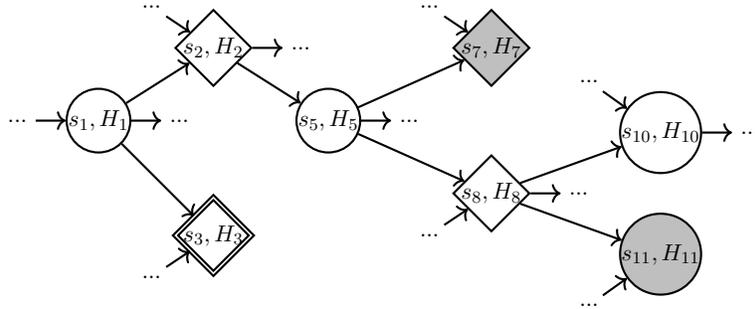
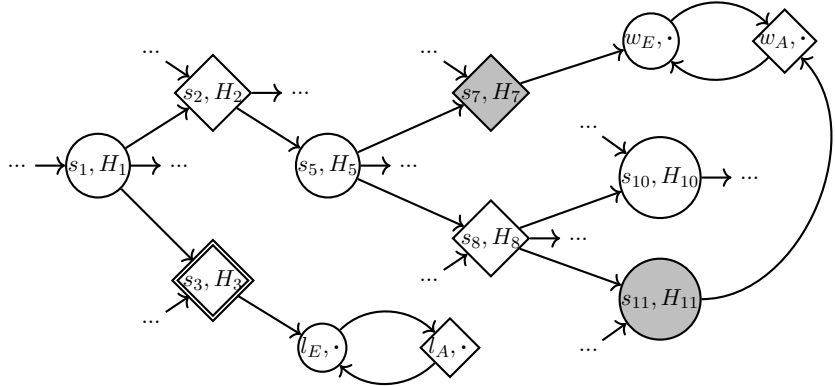

		In this step, a part of the situation graph is constructed for the set of constraints $\mathcal{C}$. Starting form the initial situation, the graph is built up by adding successors of already identified situations. Different from Definition~\ref{def:situationGraph}, we add three criteria that restrict for which situations successors are actually added to the graph:
		\begin{itemize}
			\item If a situation is an extension of a situation belonging to the winning region of a situation graph from a previous increment, outgoing transitions do not need to be considered. In fact, the situation can directly be marked as belonging to the winning region of the currently constructed situation graph.	
			\item No outgoing transitions of situations that violate any constraint of $\E$ in $\mathcal{C}$ are considered, since they are clearly not part of any winning strategy for $\E$ in the game $(A, \text{Win}, \mathcal{C})$. 
			Note that an extension of a situation that violate a constraint can again satisfy a constraint. Therefore, such extensions are considered again in later increments.
			\item Situations that violate any constraint of $\A$ are not added to the situation graph. This is not leading to any relevant restriction on the behavior of $\A$ due to the last property in Definition~\ref{def:gameWithCounstraints} and Theorem~\ref{theorem:monotony:validity}.
		\end{itemize}	
		The first criterion on not extending the situation graph from extensions of situations of the winning region of a previous increment is a key factor in reducing the required computational time and memory for incremental synthesis. While the increasing lengths of the considered counting constraints extends the lengths of the relevant histories and the number of potentially reachable situations, avoiding to construct parts of the situation graph by using knowledge retrieved in previous increments prunes the graph again. 
		Figures~\ref{subfig:sitgraph_nach_def} and \ref{subfig:sitgraph_shortened} illustrate the difference between the definition of the situation graph (Definition~\ref{def:situationGraph}) and the modified version in this step.
		
		\subsection*{Modify situation graph to incorporate the winning condition \text{Win}}
		The goal of this step is to adapt the situation graph of the previous step, such that any synthesis algorithm that can compute the winning region of the original game $(A, \text{Win})$ (without counting constraints) can be used for calculating (a part of) the winning region of the situation graph, which translate to (a part of) the winning region of $(A, \text{Win},\mathcal{C})$. To achieve this, there are three things to consider: (1) The winning condition \text{Win} needs to be extended to the situation graph, (2) information from previous increments on winning regions needs to be encoded into the game and (3) counting constraint fulfillment needs to be encoded into the graph. The latter two aspects will automatically remove all deadlocks from the situation graph. Those deadlocks are created by the two discontinuation-criteria of the previous step, but we demand deadlock-freedom for two-player games in Definition~\ref{def:twoPlayerGameGraph}.\\
		In the following, we define two sinks $W$ and $L$ that are incorporated into the situation graph for each winning condition and then go into specific adaptions for reachability as winning condition. For other winning conditions like safety, Büchi, co-Büchi and  parity, the integration of the additional sinks is a standard procedure.
		For staying in a two-player game, both sinks consists of one state controlled by $\E$ and one state controlled by $\A$ with transitions between the two states in both directions. The history-part of the states and the actions played when choosing one of the transitions are irrelevant in the following process. Denote the two states of $W$ with $(w_{E}, \boldsymbol{\cdot})$ (controlled by $\E$) and $(w_{A}, \boldsymbol{\cdot})$ (controlled by $\A$) with $\boldsymbol{\cdot}$ being a placeholder for any history and $w_E$, $w_A$ symbolizing states that are not already in the original game graph $A$. Likewise, denote the states in $L$ as  $(l_{E}, \boldsymbol{\cdot})$ and  $(l_{A}, \boldsymbol{\cdot})$.
		The sinks are integrated into the (modified) situation graph from the previous step as follows:
		\begin{itemize}
			\item Sink $W$ is reached from all situations from which player $\E$ is known to win due to already acquired knowledge. This is the case when a situation is an extension of a situation in the winning region of any previous increments, which are exactly the situations satisfying all constraints without successor in the modified situation graph from the previous step. If such a situation is controlled by $\E$, a transition to $(w_{E}, \boldsymbol{\cdot})$ is added to the situation graph. Symmetrically, if such a situation is controlled by $\A$, a transition to $(w_{A}, \boldsymbol{\cdot})$ is added.
			\item Sink $L$ is reached from all situations from which $\E$ is known to lose the game. This is the case when a counting constraints of $\E$ is violated in the situation, \ie, from all situations of the modified situation graph that have no successor and violate any constraint. From such a situation, a transition to $(l_{E}, \boldsymbol{\cdot})$ is added. Note that there are only situations controlled from $\A$ that violate any constraint left in the modified situation graph, hence the only predecessor of $(l_{A}, \boldsymbol{\cdot})$ is $(l_{E}, \boldsymbol{\cdot})$.
		\end{itemize}
		Denote the resulting graph with sinks as $Sit_{\mathcal{C}} = (\tilde{S}, s_{init},\tilde{S}_{\E}, \tilde{S}_{\A}, \tilde{\hookrightarrow})$ with $\tilde{S}$ being all situations in the modified situation graph of the previous step plus the four sink states, $s_{init}$ the initial state of the situation graph as in Definition~\ref{def:situationGraph}, $\tilde{S}_{\E}$, $\tilde{S}_{\A}$ forming a disjoint union of $\tilde{S}$ as situations controlled by $\E$ and $\A$, respectively, and $\tilde{\hookrightarrow}$ denoting the transitions in the graph with sinks.
		With a slightly abuse of notation, we call the resulting graph $Sit_{\mathcal{C}}$ with sinks again situation graph.
		Figure~\ref{subfig:sitgraph_with_sinks} illustrates the integration of the two sinks $W$ and $L$.\\
		\textbf{Reachability winning condition $\text{Win} = \mathtt{Reach}(R)$:} Compared to other winning conditions like safety, Büchi, co-Büchi or parity, adapting the winning condition and the situation graph for being solvable by an algorithm for reachability games is more complex. This is due to the fact that all of the other considered winning conditions share the \enquote{infinity aspect} with the condition of satisfying a set of window counting constraints, but for winning a play of a reachability game, it suffices to visit one of the winning states in $R$ just once. For a reachability game, the construction of $\widetilde{\text{Win}}$ as for the other winning conditions would lead to plays that visit any situation $(s,H)\in \tilde{S}$ with $s\in R$ and violating a counting constraint of $\E$ later as being in the winning region. To circumvent this issue, it is possible to play a safety game first and play the reachability game only on the winning region of the safety game:
		\begin{enumerate}
			\item Calculate the winning region $\mathcal{W}$ of $(Sit_{\mathcal{C}}, \mathtt{Safe}(\{(s,H) \in \tilde{S} \, | \, s \in S \cup \{w_E, w_A\}\}))$ with any synthesis algorithm applicable to two-player safety games. The winning region consists of the largest subgraph of $Sit_{\mathcal{C}}$ in which $\E$ can guarantee to not violate any of its constraints.
			\item Define $\tilde{R}:= \{(s,H) \in \mathcal{W} \, | \, s\in R\cup \{w_E, w_A\}\}$. With  $\widetilde{\text{Win}}:= \mathtt{Reach}(\tilde{R})$, the game $(\mathcal{W}, \widetilde{\text{Win}})$ becomes a two-player reachability game with winning plays containing at least one situation with state in $R$ and not violating any counting constraint of $\E$.
		\end{enumerate}
		
		With those modifications, the (adapted) situation graph with (adapted) winning condition becomes a game $G_{\mathcal{C}}(Sit_{\mathcal{C}}, \widetilde{\text{Win}})$ of the same type as the original game $G=(A,\text{Win})$, such that a synthesis algorithm of the readers choice can be applied to determine the winning region. Note that the construction of $\widetilde{\text{Win}}$ is in this form only possible since we only increment over counting constraints of the form $CC_{min}(\E, a, k, l)$ (with constraints of the form $CC_{max}(\E, a, k, l)$ optionally translated by using Lemma~\ref{lemma:translatonOfConstraints}). A key property that we are using here is that extensions of situations in the winning region of one increment are again in the winning region. If incrementing would go directly over $CC_{max}(\E, a, k, l)$ with increasing constraint length, this property would not hold.
		
		\subsection*{Determine the winning region of $G_{\mathcal{C}}(Sit_{\mathcal{C}}, \widetilde{\text{Win}})$}
		In this step, the winning region of $G_{\mathcal{C}}=(Sit_{\mathcal{C}}, \widetilde{Win})$ with the constructions of the previous step is calculated by any synthesis algorithm that can calculate the winning region of the original game $G=(A, \text{Win})$ without window counting constraints. The calculated winning region serves as input for following increments, especially for constructing the next situation graph. In fact, this information on the winning region is the key enabler to speed up the next increment.
		Note that the calculated winning region in this step is in general only a subset of the winning region of $(A, \text{Win}, \mathcal{C})$, as a price for not extending the situation graph in situations that are extensions of the winning region of previous increments. However, this is not a problem as long as the overall goal is to find a winning strategy for $\E$ and not the full winning region. We can just observe that $\E$ will satisfy shorter (and hence more restrictive) counting constraints than necessary when using the presented incremental approach. 
		
		\subsection*{Check if a winning strategy is already found}
		If the initial state of the adapted situation graph $Sit_{\mathcal{C}}$ is in the winning region, a set of winning strategies for $\E$ is found for the game $G_{\mathcal{C}}=(Sit_{\mathcal{C}}, \widetilde{\text{Win}})$. Any winning strategy in this game can easily be transformed to a (in general non-positional) winning strategy in $G=(A, \text{Win}, CC)$. In this case, the incrementing process stops.\\
		If the initial state of the adapted situation graph is not in the winning region and $\mathcal{C} = \mathscr{C}$ (that is, the current increment was already on the full constraint length for all constraints), no winning strategy for $\E$ in $G=(A, Win, CC)$ exists and the algorithm stops. If $\mathcal{C} = \mathscr{C}$ does not yet hold, another increment with longer constraints follows.
		
		\subsection*{Increase counting constraint lengths} 
		In this step, the lengths of counting constraints in $\mathcal{C}$ of the form $CC_{min}(\E, a, k, l)$ is increased (at most to their respective lengths in $\mathscr{C}$). Which constraint length is increased by how much is the readers choice. There is no mode for constraint length increase that is optimal for each game with window counting constraints. Two canonical implementation modes for this step are the following:
		\begin{itemize}
			\item Sequential: In this mode, constraints are extended one after the other. Choose a fixed order of the counting constraints of the form $CC_{min}(\E, a, k, l)$ in $\mathcal{C}$. Starting by the first constraint, extend the constraint length by one, if the length is not yet as in $\mathscr{C}$. Only if this upper bound is already reached, go to the next constraint and increase that one by one. 
			\item Round-robin: In this mode, the constraints are extended in turns. Again, choose a fixed order $C_1, \dots, C_m$ of the counting constraints of the form $CC_{min}(\E, a, k, l)$ in $\mathcal{C}$. In the first increment, the first constraint that is not yet completely extended is extended by one. For following increments: If in the last increment constraint $C_i$ was extended, extend now the next constraint after $C_i$ that has not yet reached its full length as in $\mathscr{C}$ (consider the ordering as circular, \ie modulo $m$). 
		\end{itemize}
		
		The updated set $\mathcal{C}$ of counting constraints together with the set of already calculated winning regions and the original serves as new input for the next synthesis increment, starting with the construction of the next situation graph.
	
	\section{Example and Experimental Results}\label{sec:experiments}
\begin{figure}
			\centering
			\begin{tikzpicture}[node distance={16mm}, thick, scale=.6] 
				\node[ego] (1) {$1$}; 
				\node[alter] (2) [right of=1] {$2$}; 
				\node[ego] (3) [above right of=2] {$3$};
				\node[alter] (4) [right of=3] {$4$};
				\node[ego] (5) [above of=4] {$5$};
				\node[alter] (6) [above of=3] {$6$};
				\node[ego] (7) [below right of=2] {$7$};
				\node[alter] (8) [right of=7] {$8$};
				\node[ego] (9) [right of=8] {$9$};
				\node[alter] (10) [right of=5] {$10$};
				
				\draw[->] (-1,0) -- (1); 
				\draw[->] (1) --  (2); 
				\draw[->] (2) --  (3); 
				\draw[->] (3) --  (4);
				\draw[->] (4) --  (5); 
				\draw[->] (5) -- node[midway, above] {$a$} (6); 
				\draw[->] (6) --  (3);
				\draw[->] (2) -- (7); 
				\draw[->] (7) -- node[midway, above] {$a$} (8); 
				\draw[->] (8) to [out=45,in=135]  (9); 
				\draw[->] (9) to [out=225,in=315]  node[midway, below] {$a$} (8); 
				\draw[->] (9) -- (10); 
				\draw[->] (10) --  (5); 
			\end{tikzpicture} 
			\caption{Two-player game graph $A$ for a safety game. States represented as circles are controlled by $\E$, diamond-shaped states are controlled by $\A$. $\Sigma_{E}=\{a\}$ and outgoing transitions of states controlled by $\E$ without label mean that $\E$ plays $\neg a$ in those moves. The played letters of $\A$ are not important in this example, hence labels of outgoing transitions of $\A$-controlled states are omitted. To keep the example simple, we consider a safety game without unsafe states, that is, $\text{Win} = \mathtt{Safe(R)}$ for $R$ containing all states of $A$.  $\E$ shall fulfill the counting constraint $C=CC_{min}(\E, a,1,7)$ ($\E$ plays $a$ at least one time in 7 turns), making $G=(A, \text{Win}, \{C\})$ a game with counting constraints.}
			\label{fig:smallGameIterationsMatter}
		\end{figure}
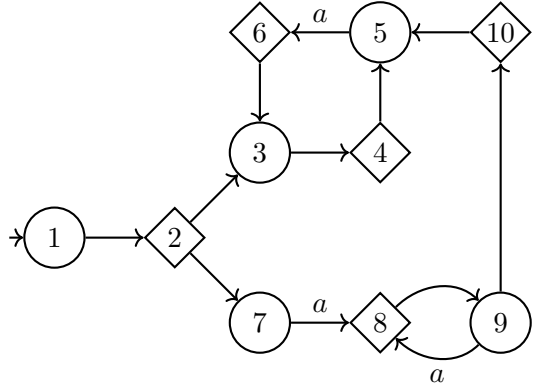

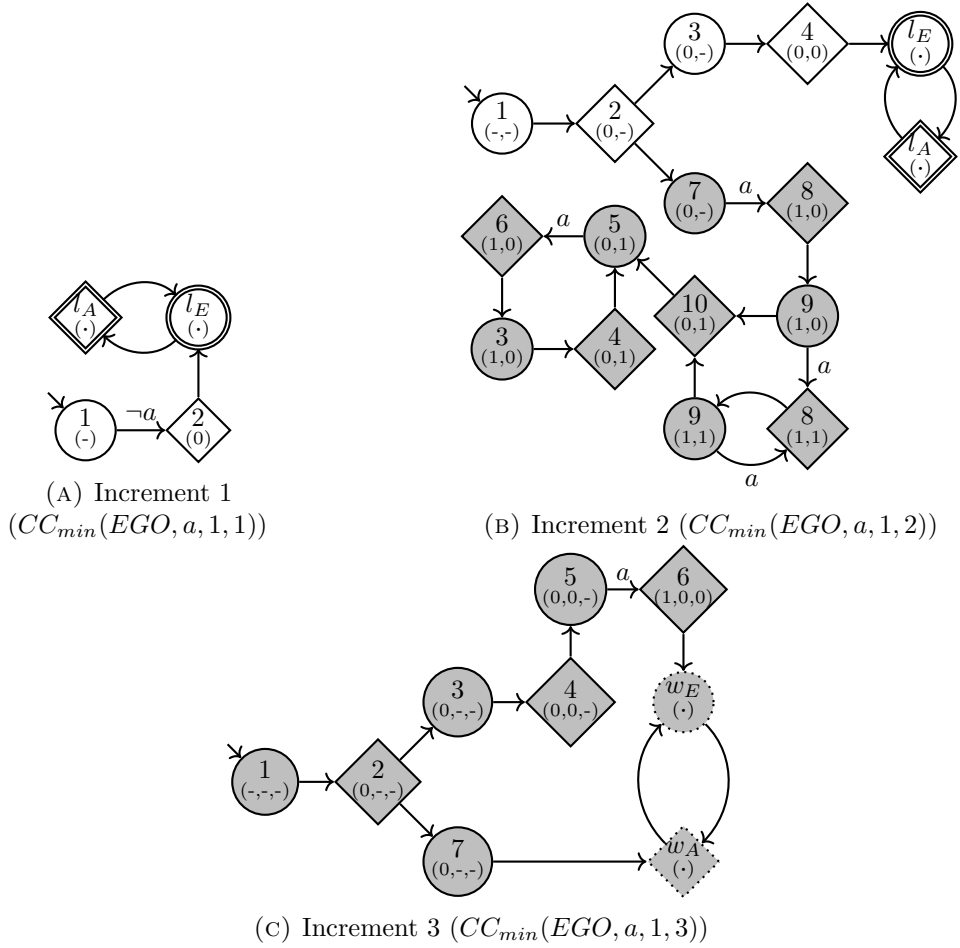
\begin{figure}
			\centering
			\small
			\begin{subfigure}[b]{.4\textwidth}
				\centering
				\begin{tikzpicture}[node distance={15mm}, thick, scale=1.0, font=\small] 
					\node[ego] (1) {\Situation{1}{\none}}; 
					\node[alter] (2) [right of=1] {\Situation{2}{0}}; 
					
					\node[ego, double] (lE) [above of= 2] {\Situation{l_E}{\boldsymbol{\cdot}}};
					\node[alter, double] (lA) [above of=1] {\Situation{l_A}{\boldsymbol{\cdot}}};
					
					\draw[->] (-0.5,0.5) -- (1); 
					\draw[->] (1) -- node[midway, above] {$\neg a$} (2);
					
					\draw[->] (lA) to [out=45,in=135]  (lE); 
					\draw[->] (lE) to [out=225,in=315]  (lA); 
					\draw[->] (2) -- (lE);
				\end{tikzpicture}   
				\caption{Increment 1\\ ($CC_{min}(\E, a,1,1)$).}
				\label{fig:iteration1}                                     
			\end{subfigure}%
			\begin{subfigure}[b]{.6\textwidth}
				\centering
				\begin{tikzpicture}[node distance={15mm}, thick, scale=1.0, font=\small, align=center] 
					\node[ego] (12) {\Situation{1}{\none,\none}}; 
					\node[alter] (22) [right of=12] {\Situation{2}{0,\none}}; 
					\node[ego] (321) [above right of=22] {\Situation{3}{0,\none}};
					\node[ego, winnable] (72) [below right of=22] {\Situation{7}{0,\none}};
					\node[alter] (421) [right of=321] {\Situation{4}{0,0}};
					\node[alter, winnable] (821) [right of=72] {\Situation{8}{1,0}};
					\node[ego, winnable] (921) [below of=821] {\Situation{9}{1,0}};
					\node[alter, winnable] (822) [below of=921] {\Situation{8}{1,1}};
					\node[ego, winnable] (922) [left of=822] {\Situation{9}{1,1}};
					\node[alter, winnable] (102) [left of=921] {\Situation{10}{0,1}};
					\node[ego,winnable] (52) [above left of=102] {\Situation{5}{0,1}};
					\node[alter, winnable] (62) [left of=52] {\Situation{6}{1,0}};
					\node[ego, winnable] (322) [below of=62] {\Situation{3}{1,0}};
					\node[alter, winnable] (422) [right of=322] {\Situation{4}{0,1}};
					
					\node[ego, double] (lE) [right of= 421] {\Situation{l_E}{\boldsymbol{\cdot}}};
					\node[alter, double] (lA) [below of=lE] {\Situation{l_A}{\boldsymbol{\cdot}}};
					
					\draw[->] (-0.5,0.5) -- (12); 
					\draw[->] (12) --  (22); 
					\draw[->] (22) -- (321); 
					\draw[->] (22) -- (72);
					\draw[->] (321) -- (421); 
					\draw[->] (72) -- node[midway, above] {$a$} (821);
					\draw[->] (821) -- (921);
					\draw[->] (921) -- node[midway, right] {$a$} (822);
					\draw[->] (922) --  (102);
					\draw[->] (921) -- (102);
					\draw[->] (102) --  (52);
					\draw[->] (52) -- node[midway, above] {$a$} (62);
					\draw[->] (62) --  (322);
					\draw[->] (322) --  (422);
					\draw[->] (422) --  (52);
					\draw[->] (822) to [out=135,in=45]  (922); 
					\draw[->] (922) to [out=315,in=225]  node[midway, below] {$a$} (822); 
					
					\draw[->] (lA) to [out=135,in=225]  (lE); 
					\draw[->] (lE) to [out=-45,in=45] (lA); 
					\draw[->] (421) -- (lE);
				\end{tikzpicture}
				\caption{Increment 2 ($CC_{min}(\E, a,1,2)$).}
				\label{fig:iteration2}
			\end{subfigure}
			\begin{subfigure}[b]{0.5\textwidth}
				\centering
				\begin{tikzpicture}[node distance={15mm}, thick, scale=1] 
					\node[ego, winnable] (13) {\Situation{1}{\none,\none,\none}}; 
					\node[alter, winnable, ,  inner sep=-1pt] (23) [right of=13] {\Situation{2}{0,\none,\none}}; 
					\node[ego, winnable] (73) [below right of=23] {\Situation{7}{0,\none,\none}}; 
					\node[ego, winnable] (33) [above right of=23] {\Situation{3}{0,\none,\none}}; 
					\node[alter, winnable, ,  inner sep=-1pt] (43) [right of=33] {\Situation{4}{0,0,\none}};
					\node[ego, winnable] (53) [above of=43] {\Situation{5}{0,0,\none}};  
					\node[alter, winnable,  inner sep=-1.5pt] (63) [right of=53] {\Situation{6}{1,0,0}};
					
					\node[ego, dotted, winnable] (wE) [right of= 43] {\Situation{w_E}{\boldsymbol{\cdot}}};
					\node (c) [draw=none, below left of=wE]{};
					\node[alter, dotted, winnable] (wA) [below right of=c] {\Situation{w_A}{\boldsymbol{\cdot}}};
					
					\draw[->] (-0.5,0.5) -- (13);
					\draw[->] (13) --  (23); 
					\draw[->] (23) --  (73); 
					\draw[->] (23) --  (33); 
					\draw[->] (33) --  (43); 
					\draw[->] (43) --  (53); 
					\draw[->] (53) -- node[midway, above] {$a$} (63); 
					
					\draw[->] (wA) to [out=135,in=225]  (wE); 
					\draw[->] (wE) to [out=-45,in=45]  (wA); 
					\draw[->] (63) -- (wE);
					\draw[->] (73) -- (wA);
				\end{tikzpicture}
				\caption{Increment 3 ($CC_{min}(\E, a,1,3)$).}
				\label{fig:iteration3}
			\end{subfigure}
			\caption{Situation graphs for the game in \autoref{fig:smallGameIterationsMatter} with incrementing over $CC_{min}(\E, a,1,7)$. Situations belonging to the winning region are marked in gray. Situations of the losing sink $L$ are marked double-framed and situations of the winning sink $W$ with dotted lines. Each of the showed situations (nodes in the graphs) consist of a state of the original game graph in Figure~\ref{fig:smallGameIterationsMatter} and a history on how the situation was reached. For example, the history in situation $(6, (1,0,0))$ (upper right corner of Figure~\ref{fig:iteration3}) encodes that $\E$ played $a$ in its last move, but not in the second and third to last move. More than three increments are not necessary to wind a winning strategy for $\E$, since there already is a winning strategy for $\E$ in the third increment.}
			\label{fig:situationGraphs}
		\end{figure}

		For illustrating the synthesis algorithm, we consider the game graph $A$ in \autoref{fig:smallGameIterationsMatter} as small example before reporting on larger experiments for showing the potential of incremental synthesis with counting constraints. 
		Let $CC_{min}(\E, a,1,7)$ be a counting constraint that $\E$ needs to satisfy. For the sake of keeping the example small, we pass on more counting constraints, only focus on the action \enquote{$a$} of $\E$ and use a simple safety winning condition by marking all states as safe, that is $Win = \mathtt{Safe}(S)$ with $S$ being all states in the game graph $A$. Denote the game with $G = (A, \text{Win}, \mathcal{C})$ with $\mathcal{C} = \{CC_{min}(\E, a,1,7)\}$.\\
		Following the process sketched in Figure~\ref{fig:overview_iteration_min}, the first increment is using the counting constraint $CC_{min}(\E, a,1,1)$ (\enquote{$\E$ plays $a$ at least in one of 1 turns}). In the initial state $1$, $\E$ is only allowed to play $\neg a$, hence $\E$ violates this constraint already with its first (and only) turn. Consequently, the only considered successor of the reached situation is the losing sink $L$ and the situation graph remains rather small, as depicted in Figure~\ref{fig:iteration1}. Taking this situation graph as new game graph and winning condition $\widetilde{\text{Win}} = \mathtt{Safe(\{(1,(-)), (2,(0))\})}$ yields an empty winning region. For the next increment, the length of the counting constraint is increased by one, that is,  $\{CC_{min}(\E, a,1,2)\}$ is taken. The resulting situation graph is shown in Figure~\ref{fig:iteration2}. Considering all of the included situations except for the two losing sink states as safe, the winning region of the resulting safety game now includes a subgraph with 10 situations and 12 transitions (marked in gray in the figure). Since the initial situation is not part of the winning region, no winning strategy for $\E$ is found yet and the counting constraint is increased to $CC_{min}(\E, a,1,3)$. The resulting situation graph is shown in Figure~\ref{fig:iteration3}. Note that this graph is rather small, since successors of situation extensions of the previous winning region do not need to be considered, but the extensions directly go into the winning sink. For example, the situation $(7,(0,-,-))$ is an extension of the situation $(7,(0,-))$, which is in the winning region of the second increment, hence the only successor of $(7,(0,-,-))$ is the winning sink. The same holds for situation $(6, (1,0,0))$ as extension of $(6, (1,0))$. With all situations in the third situation graph marked as safe (since no counting constraint violations are included), the whole graph becomes the winning region. Hence, the initial state is also winnable and a winning strategy for $\E$ is found and no further increment (with potentially larger situation graphs) is needed.
		
		Please note that the focus on the example is to show how the situation graph evolves over multiple increments, illustrating the benefit of the incremental approach. However, the example is too small to actually be significantly more efficient than synthesizing a winning strategy without increments. We close this gap now by presenting analysis results for larger games.
		
		For this, an explicit state implementation of the approach (Figure~\ref{fig:overview_iteration_min}) in Python is used with the following implementation details.
		\begin{itemize}
			\item All constraints on $\E$ are translated to counting constraints of the form $CC_{min}(\E,$\linebreak $a,k,l)$.
			\item In the initialization step, all constraints on $\E$ are set to the minimal possible constraint length, that is, to $CC_{min}(\E, a,k,k)$.
			\item All experiments are done on safety games, hence only a backwards coloring algorithm for determining the winning region of a safety game is implemented.
			\item For increasing the counting constraint length for the next iteration, we implemented the sequential and the round-robin mode explained above. Each game is played in each of the available modes.
		\end{itemize} 
		As comparison, the same base implementation is used, but the first (and only) increment is done directly with the full length of all counting constraints, which translates to synthesis without incrementing and direct encoding of the required memory for the full constraint length into the game graph.
		Table~\ref{tab:analysisResults} summarizes the performance results of the presented incremental synthesis approach. Each game was solved three times: first in sequential mode for increasing the constraints, second in round-robin mode and the third time without incrementing (by going directly to the full constraint lengths). For each of those runs of the synthesis algorithms, the required computational time and the number of states in the last considered situation graph is presented in the table. Additionally, it is noted how many constraints on $\E$ were given and if a winning strategy for $\E$ exists. A question mark \enquote{?} in the table stands for data that was not recorded due to hardware limitations (thrashing due to reached memory limits; in fact, this phenomenon could also have lead to the high computational time in experiment number 2 without incrementing).
		
		\begin{table}[htbp]
			\centering
			\caption{Runtime comparison of synthesis with and without incrementing}
			\label{tab:analysisResults}
			\begin{tabular}{lllcccccc}
				\toprule
				\multicolumn{3}{c}{} & \multicolumn{4}{c}{with incrementing} & \multicolumn{2}{c}{without incrementing} \\
				\cmidrule(lr){4-7} \cmidrule(lr){8-9}
				\multicolumn{3}{c}{} & \multicolumn{2}{c}{sequential} & \multicolumn{2}{c}{round-robin} & \multicolumn{2}{c}{} \\
				\midrule
				No. & Win? & $|\mathcal{C}|$ & $|$states$|$ & time [sec] & $|$states$|$ & time [sec] & $|$states$|$ & time [sec] \\
				\midrule
				1 & Yes & 1 & 284,636 & 37 & 284,636 & 42 & 3,692,986 & 316 \\ 
				2 & Yes & 2 & 90,842 & 11 & 9,742 & 3 & 6,097,599 & 6102 \\ 
				3 & No & 3 & 46 & 45 & 46 & 56 & 1,723,345 & 195 \\ 
				4 & Yes & 4 & 1,129,237 & 552 & 146,769 & 36 & ? & ? \\ 
				5 & Yes & 2 & 537,970 & 111 & 152,882 & 51 & ? & ? \\ 
				6 & Yes & 1 & 6,297 & 0.4 & 6,297 & 0.4 & 3,031,519 & 579 \\
				7 & Yes & 2 & 1,990,179 & 709 & 1,538,446 & 626 & 3,351,204 & 544 \\ 
				\bottomrule
			\end{tabular}
		\end{table}
		
		The experiments were manually constructed and are not modeling real use cases. Some of the experiments were drawn as grid games that were automatically translated to game graphs. Those grid games were loosely motivated by the challenge of navigating a robot on a factory floor, while the robot should visit predefined areas on a regular basis (defined as counting constraints on $\E$) and some tiles of the grid might be temporarily blocked by the environment (think e.g. of a fork lifter). To give an impression on the experiments, Figure~\ref{fig:exampleGame5} shows the grid game on which the game graph for Experiment 5 in Table~\ref{tab:analysisResults} is based on. 
		
		\begin{figure}
			\centering
			\begin{tikzpicture}
				\draw[thick] (0,0) grid (3,3);
				
				\node at (0.2, 2.8) {\textbf{m}};
				\node at (2.2, 2.8) {\textbf{c}};
				
				\fill[gray!30] (1, 1) rectangle (2, 2);
				
				\draw[thick] (0.5, 2.1) -- (0.2, 2.4) -- (0.5, 2.7) -- (0.8, 2.4) -- cycle; 
				\draw[thick] (1.5, 1.4) circle (0.3); 
			\end{tikzpicture}
			\caption{Grid game illustrating experiment 5. 
				The circle and the diamond indicate the (initial) positions of $\E$ and $\A$, respectively. The \textbf{m} (machine) and \textbf{c} (charging station) in the upper left and right corners represent target cells of $\E$: $\E$ is required to fulfill the counting constraints \enquote{$\E$ plays \textbf{c} at least 1 times out of 13 of its own turns.} and \enquote{$\E$ plays \textbf{m} at least 1 times out of 13 of its own turns.}, meaning that $\E$ shall visit the charging station and the machine \textbf{c} at least once in 13 turns, respectively. $\A$ has the constraints \enquote{"$\A$ plays circle at least 1 times out of 1 of its own turns."} (meaning that $\A$ is not allowed to visit the shaded cell in the middle) and \enquote{$\A$ plays (not idle) at least 2 times out of 4 of its own turns.} (meaning: $\A$ is only allowed to stay at the same cell instead of moving to another cell in at most 2 out of 4 turns). The game is played as a safety game with crashes of the two players (both players on the same cell) as unsafe states.}
			\label{fig:exampleGame5}
		\end{figure}
		
		The realized comparisons show the great potential of successively enlarging counting constraints, allowing for incomplete graph constructions due to retrieved information on already winnable states of prior graphs instead of encoding the full constraints directly in a graph for strategy synthesis. In all but one of the experiments, the incremental synthesis required significantly less time and memory. Note that the number of states in Table~\ref{tab:analysisResults} refers to the number of states in the last constructed situation graph. If incrementing is used, additional memory for the states of the already calculated winning regions of previous increments is required. It suffices to remember a minimal set $M$ of states, such that each state of any of the already calculated winning regions is an extension of a state in the minimal set $M$. The number of states in $M$ can be larger than the next situation graph (as can be seen in the example above and this is also the case for experiment 3 in the table). However, the set $M$ is (heuristically) much smaller than the situation graph without incrementing. For example, experiment 3 in Table~\ref{tab:analysisResults} required to remember 68,937 states of winning regions for sequential increments and 8,401 states for round-robin increments, which is orders of magnitude smaller than the 1,723,345 states in the situation graph without incrementing. 
		As can be seen in Experiment 7, incremental synthesis is not in general better for games with counting constraints, but can be outperformed by classical synthesis without iteration for some games. This can for example happen if the full length of the counting constraints is required to actually find a winning strategy for $\E$ and the winning region in former increments stays empty. In such cases, the incremental approach only adds a non-beneficial overhead and the only learned additional information is that the full constraint length is actually needed, which does not contribute to runtime savings.

		\section{Discussion and Future Work} \label{sec:discussion}
		
		The exploitation of the monotonicity property inherent in counting constraints for incremental synthesis has demonstrated promising outcomes, indicating the potential for time- and memory-efficient computation of controllers for reactive systems. The current investigation aimed to explore the broader applicability of incremental synthesis utilizing counting constraints, an objective that has been achieved. However, certain challenges and considerations in the chosen game setting shall be discussed in the following, paving the way for future research directions.
		
\begin{figure}
			\centering
			\begin{tikzpicture}[node distance={16mm}, thick, scale=0.8] 
				\node[ego] (1) {$1$}; 
				\node[alter] (2) [right of=1] {$2$}; 
				\node[ego] (3) [below of=2] {$3$};
				\node[ego] (4) [right of=2] {$4$};
				\node[alter] (5) [right of=4] {$5$};
				\node[ego] (6) [below of=5] {$6$};
				\node[alter] (7) [below of=4] {$7$};
				
				\draw[->] (-1,0) -- (1); 
				\draw[->] (1) -- node[midway, above] {$a$} (2); 
				\draw[->] (2) to [out=225,in=135]  node[midway, left] {$b$} (3); 
				\draw[->] (3) to [out=45,in=315]  node[midway, right] {$a$} (2); 
				\draw[->] (2) -- node[midway, above] {$\neg b$} (4); 
				\draw[->] (4) -- node[midway, above] {$a$} (5); 
				\draw[->] (5) -- node[midway, right] {$\neg b$} (6); 
				\draw[->] (6) -- node[midway, above] {$a$} (7); 
				\draw[->] (7) -- node[midway, left] {$b$} (4); 
			\end{tikzpicture} 
			\caption{$\A$ can always fulfill the counting constraint $CC_{min}(\A, b, 1, 3)$, but can run into a violation for $CC_{min}(\A, b, 1, 2)$.}
			\label{fig:invalidGame}
		\end{figure}
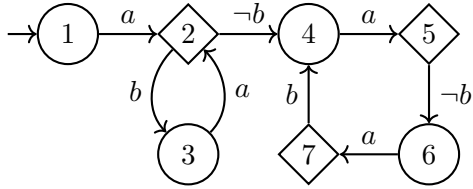

		\textbf{Towards cooperative games}: 
		The presented approach did not consider incrementing over constraints of $\A$, but handled them as being fixed. The rationale for this lies in the property \enquote{$\A$ cannot be forced into constraint violations} of games with counting constraints (Definition~\ref{def:gameWithCounstraints}). 
		In general, a game with counting constraints may satisfy the requirement of $\A$ always being able to adhere to its counting constraints, only to find the requirement violated for the game with a modified counting constraint as used in the increments. An illustrative example is provided in \autoref{fig:invalidGame}. $\A$ has the counting constraint  $CC_{min}(\A, b, 1, 3)$, i.e.\ $\A$ plays $b$ at least once in three of its turns. Recall that \enquote{$\A$ cannot be forced into constraint violations} is defined  as $\A$ is able to enlarge each prefix that satisfies the constraint such that the resulting prefix is also satisfying the constraint. This property is fulfilled when considering the game graph and the constraints $CC_{min}(\A, b, 1, 1)$ or $CC_{min}(\A, b, 1, 3)$. However, it is violated for $CC_{min}(\A, b, 2, 3)$, since the prefix $(1, a, 2, \neg b, 4, a, 5)$ satisfies the constraint\footnote{$\A$ could play $b$ forever to complete the prefix to an infinite play that satisfies the constraint. This play is not in $G$, but nonetheless is sufficient according to the definition of a prefix satisfying  a constraint in \autoref{def:windowCountingConstraints}.}, but there is no possibility for $\A$ to still satisfy the constraint with the next turn. Since the definition of a winning strategy relies on the game property of $\A$ not being forceable into counting constraint violations, \autoref{theorem:monotony} cannot be extended to increments over $\A$-constraints. However, such an extension would offer additional potential for more efficient synthesis algorithms.
		
		We plan to approach this problem by leaving the zero-sum setting. The environment wins if it has a strategy that guarantees to satisfy all of its counting constraints. In particular, it is possible that the environment violates a constraint and loses. The envisioned game setting shall avoid the well-known problem of $\E$ winning only by falsifying the assumptions in form of counting constraints on $\A$. Instead, $\E$ shall support $\A$ in satisfying all constraints as long as this does not compromise the adherence to own constraints. This leads us in the direction of searching for strategy profiles with certain properties as synthesis results instead of winning strategies only for $\E$ with the exact profile properties yet to be determined. It can be foreseen that this setting requires more synchronization between the players than that presented by a zero-sum setting, in which $\A$ does not even need knowledge on counting constraints of $\E$.
		
		\textbf{Extension of counting constraint types:} It is worth to consider additional specification patterns with similar monotonicity properties as the presented counting constraints. For instance, a pattern like \enquote{if $x$ is played, $\E$ plays $y$ after at most $k$ turns} is frequently used as specification. Satisfying such a specification becomes easier for larger $k$. 
		In terms of an incremental algorithm, this means that states of the situation graph for some increments are winnable, if the related state is winnable in an earlier increment. 
		The identification of additional counting constraints and the adaption of the incremental strategy synthesis algorithm to such constraints extends the applicability of the approach to more systems.
		
		\textbf{Symbolic representation:} The presented synthesis approach uses an explicit representation of states in the situation graph as arena. However, symbolic synthesis proves to be significantly more efficient than explicit synthesis algorithms for many (but not all) applications \cite{Finkbeiner2016}. Since the presented approach already has similarities to antichains and the states of the arena have a special structure (representing a snippet of the history of a play), we expect that the approach can be transformed into a symbolic algorithm. We plan to investigate a symbolic version of the algorithm and to compare its performance with its explicit version.
		
		\textbf{Combining the presented work with on-the-fly synthesis algorithms:} 
		Other synthesis algorithms pursuing an on-the-fly forward expansion of a game graph would be a natural point of comparison to the presented approach. In order to apply these algorithms in the setting as in this article, they would however have to be applied to the full game graph with memory for the counting constraints, which they would then expand on-the-fly. We would argue that this is orthogonal to our incremental expansion of the game graph by successively refined counting constraints, and that these on-the-fly algorithms could consequently be combined with our approach. An experimental evaluation however requires substantial implementation effort and is considered as future work. We plan to investigate if the general ideas in the submission can be used for guiding the state space exploration of on-the-fly-methods when facing counting constraints (or other constraints with some monotonicity property) as part of the specification.
		
		\textbf{Symmetric synthesis:} In many application scenarios, the considered system is a system of systems, whose individual systems are identical in their hardware and run in the same context. From the development, production and maintenance point of view, it is preferable to also have the same software and in particular the same controller for all of those systems, as it is typically realized in industrial applications. An example is a fleet of transport robots that take over logistic tasks on a factory floor. 
		Some synthesis approaches that consider such symmetric synthesis (\eg, \cite{EhlersFinkbeiner2017}, \cite{CaltaShkatov2011}, \cite{RosaCuryBaldissera2024}) exist, but to the best of our knowledge, there still remain open gaps.
		We are especially interested in such symmetric setting because (1) it is of high relevance for the same application domain where the problem of synthesis with counting constraints arose (fleet of transport robots) and \newline (2) because the requirement of all symmetric systems behaving the same when facing the same situation induces again some structure in the specification of the systems that we want to explore, similar to the motivation behind investigating counting constraints.
		
		\section{Conclusion}\label{sec:conclusion}
		Synthesis algorithms for reactive systems are promising tools for various engineering tasks, most prominently for the creation of correct-by-construction controllers and for checking the feasibility of specifications. The efficiency of such algorithms is a challenge for getting synthesis into application, since the translation of the system specification into an automaton that is suitable for synthesis is costly in terms of memory and computational time. 
		The exploitation of specific properties in the specification can help to overcome this challenge. 
		In this article, we have shown the potential of incremental synthesis algorithms for specifications with monotonicity properties as for the presented counting constraints. 
		With each increment, the automaton encoding the specification is becoming larger.
		The key idea is to gather information in each increment that can be used in the next increment to reduce the size of the automaton. The precise nature of this information depends on the considered specification. 
		We have shown an incremental algorithm for a set of counting constraints of the form \enquote{the system does a specific move $m$ at least/at most in $k$ turns out of $l$}, in which information on winnable states of the automaton in one increment can be used to determine which parts of the automaton for the next increment do not need to be constructed. The approach allows to tackle games with various winning conditions and the integration of specialized synthesis algorithms for calculating the winning regions of intermediate games occurring in the incrementing process. In the majority of the presented experiments, the incremental approach requires significantly less memory and computational time than direct synthesis with full specification translation into one automaton. 
		As future work, we plan to extend the incremental synthesis in five dimensions: (1) Consideration of more cooperative behavior between system and its environment instead of a purely adversarial setting, (2) identification of new specification types with monotonicity properties that can be exploited via incremental synthesis, (3) transformation of the synthesis approach to a symbolic synthesis algorithm, (4) investigation if on-the-fly-methods can benefit from the structural properties inherent in games with counting constraints and (5) consideration of systems of systems, in which identical systems follow the same strategy.

		\bibliographystyle{alphaurl}
		\bibliography{references}

	\end{document}